\documentclass[reprint,onecolumn,eqsecnum,showpacs,nofootinbib,aps,prd,notitlepage,%
amsmath,amssymb,superscriptaddress]{revtex4-1}

\usepackage[dvips]{graphicx}
\usepackage{amsmath,amssymb}
\usepackage{amsfonts,amssymb,mathrsfs}

\usepackage{calc,enumerate}

\usepackage[usenames,dvipsnames]{color}

\usepackage{hyperref}

\usepackage{subfigure}

\input cyracc.def
\newfam\cyrfam
\font\tencyr=wncyr10
\font\sevencyr=wncyr7
\font\fivecyr=wncyr5
\def\cyr{\fam\cyrfam\tencyr\cyracc}
\textfont\cyrfam=\tencyr \scriptfont\cyrfam=\sevencyr
\scriptscriptfont\cyrfam=\fivecyr
\newcommand{\ppi}{{\cyr p}}

\DeclareMathOperator*{\Pl}{Pl}
\newcommand{\lPl}{\ell_{\Pl}}
\newcommand{\rd}{{\rm d}}
\newcommand{\re}{\mathbb{R}}

\newcommand{\Dom}{\mathcal{D}}
\newcommand{\Hil}{\mathcal{H}}
\newcommand{\gr}{{\rm gr}}

\newcommand{\Id}{\openone}

\newcommand{\ket}[1]{|{#1}\rangle}

\newcommand{\braket}[2]{\langle{#1}|{#2}\rangle}

\def\be{\begin{equation}}
\def\ee{\end{equation}}
\def\nn{\nonumber}
\def\f{\frac}

\def\sgn{{\rm sgn}}
\def\pl{{\rm Pl}}
\def\lp{\ell_\pl}
\def\mC{\mathcal{C}}
\def\mH{\mathcal{H}}
\def\mL{\mathcal{L}}

\def\mV{\mathcal{V}}

\def\h{\hat}

\def\ul{\underline}
\def\wh{\widehat}
\def\dd{{\rm d}}

\def\del{\partial}
\def\al{\alpha}
\def\de{\delta}
\def\ga{\gamma}

\def\om{\omega}

\def\vp{\varphi}
\def\aA{\left( {}^\alpha A \right)}
\def\aF{\left( {}^\alpha F \right)}

\def\aL{{}^\alpha \mL}
\def\aPi{\left( {}^\alpha \Pi \right)}
\def\aH{{}^\alpha \mH}
\def\Ag{A_\gamma}
\def\Pg{\Pi_\gamma}
\def\wPg{\wh \Pi_\gamma}
\def\wAg{\wh A_\gamma}
\def\uwPg{\wh{\ul\Pi}_\gamma}

\begin{document}

\title{Loop Quantum Cosmology of a Radiation-Dominated\\ Flat FLRW Universe}

\author{Tomasz \surname{Paw{\l}owski}} 
  \email{tomasz.pawlowski@unab.cl}
  \affiliation{Departamento de Ciencias F\'isicas, Facultad de Ciencias Exactas,\\ 
    Universidad Andres Bello, Av.~Rep\'ublica 220,  Santiago 8370134, Chile.}
  \affiliation{Wydzia{\l} Fizyki,
    Uniwersytet Warszawski, Ho\.{z}a 69, 00-681 Warszawa, Poland}

\author{Roberto \surname{Pierini}} 
  \email{roberto.pierini@unicam.it}
  \affiliation{Aix Marseille Universit\'e, CNRS,
    CPT, UMR 7332, 13288 Marseille, France}
  \affiliation{Universit\'e de Toulon, CNRS, CPT,
    UMR 7332, 83957 La Garde, France}
  \affiliation{School of Science and Technology, University of Camerino,
    I-62032 Camerino (MC), Italy}
  \affiliation{INFN Sezione di Perugia, 06123 Perugia, Italy}

\author{Edward \surname{Wilson-Ewing}} 
  \email{wilson-ewing@phys.lsu.edu}
  \affiliation{Aix Marseille Universit\'e, CNRS,
    CPT, UMR 7332, 13288 Marseille, France}
  \affiliation{Universit\'e de Toulon, CNRS, CPT,
    UMR 7332, 83957 La Garde, France}
  \affiliation{Department of Physics \& Astronomy, Louisiana State University,
    Baton Rouge, LA 70803-4001, USA}

\pacs{98.80.Qc, 04.60.Pp}

\begin{abstract}
	We study the loop quantum cosmology of a flat Friedmann-Lema\^itre-Robertson-Walker 
	space-time with a Maxwell field. We show that many of the qualitative properties 
	derived for the case of a massless scalar field also hold for a Maxwell field. In
	particular, the big-bang singularity is	replaced by a quantum bounce, and the
	operator corresponding to the matter energy	density	is bounded above by the same
	critical energy density. We also numerically study the evolution of wave functions
	that are sharply peaked in the low energy regime, and derive effective equations
	which very closely approximate the full quantum dynamics of sharply peaked
	states at all times, including the near-bounce epoch. In the process, the
	analytical and numerical methods originally used to study the dynamics in
	LQC for the case of a massless scalar field are substantially improved to handle 
	the difficulties (that generically arise for matter content other than a 
	massless scalar	field) related to the presence of a Maxwell field.
\end{abstract}

\maketitle

\section{Introduction}
\label{s.intro}

In loop quantum cosmology (LQC) \cite{as-rev, *bcmb-rev, *ac-rev, *bojo-liv}, cosmological
models are quantized in a non-perturbative manner using the basic operators and following
the methods of loop quantum gravity (LQG). The first cosmologies to be studied were the
homogeneous and isotropic Friedmann-Lema\^itre-Robertson-Walker (FLRW) space-times
with a massless scalar field,
where it was initially shown that the quantum equations of motion do not break down
at the big-bang singularity \cite{boj-flrw}, and then that the singularity is replaced by
a quantum gravity ``bounce'' that connects a pre-bounce contracting FLRW space-time to
a post-bounce expanding FLRW space-time \cite{aps-imp, acs}.

These results have since been generalized to include FLRW space-times that 
allow non-zero spatial curvature \cite{apsv-spher, skl-spher, *v-open, *s-open},
a non-zero cosmological constant \cite{bp-negL, *pa-posL} or have pressureless
dust as the matter content \cite{hp-dust-LQC} (the latter as a test for the full 
LQG framework proposed in \cite{hp-dust-LQG} where an irrotational pressureless
dust field acts as a clock%
\footnote{The deparametrization of general relativity with respect to
	the irrotational dust field leads to a formulation of LQG
	with a true Hamiltonian for the gravitational and non-dust
	matter degrees of freedom, and circumvents a series of
	technical obstacles in completing the quatization program.
	See also \cite{s-dust}.}).
In all of these cases the big-bang singularity has also been shown to
be replaced by a bounce. Space-times that allow anisotropies \cite{awe-b1, mbgmmwe, awe-b2,
*we-b9, *swe} and inhomogeneities (using a hybrid quantization procedure) \cite{gmbmm,*mbmmwe,
mbmdbmm} have also been studied in LQC; in the Bianchi and Gowdy models, the classical
singularity is resolved as the singular states decouple from the non-singular states under
the quantum dynamics. It is generally expected that the big-bang singularity is replaced by
a bounce in this setting as well (see, e.g., studies of the effective equations for the LQC
of the Bianchi I \cite{gs-b1} and Gowdy \cite{bmp-geff-ftc,*bmp-geff-det} space-times),
but this has not yet been shown as the full quantum dynamics have not yet been investigated.
Indeed, dynamical studies of inhomogeneous cosmologies in LQC rely strongly
on the extrapolation of the properties of systems studied at a genuinely quantum level, 
in particular the preservation of semi-classicality and the validity of the semi-classical
effective dynamics as a good approximation to the quantum dyanmics.

Most recently cosmological perturbations have also been studied in LQC, first from an
effective theory standpoint \cite{bhks,*we-pert1,*cmbg} and then in quantum treatments 
following various approaches: a ``hybrid'' quantization (i.e., the LQC of
the homogeneous background and a Fock quantization of the perturbative degrees of freedom)
\cite{fmo,*aan-pert2} (see also \cite{dlp-qft}) and by the LQC treatment of a discretization
of the flat FLRW space-time with scalar perturbations \cite{we-pert2}. These works have allowed
the study of the dynamics of cosmological perturbations in the Planck regime in some of the most
interesting cosmological scenarios ---those that generate a scale-invariant power spectrum
of scalar perturbations--- namely inflation \cite{bct,*lcb,*aan-pert3}, the matter bounce
\cite{we-mb,*cwe} and the ekpyrotic universe \cite{we-ekp}.

In LQC, the main research effort is focused on investigating the loop quantum 
geometry effects, arising from a quantization which differs from the standard 
Wheeler-DeWitt one. The matter sector is usually dealt with in a perfunctory manner, 
although there do exist some studies on the polymeric matter sector in the literature
\cite{ldd-scalar1,*ldd-scalar2,kp-polysc} (including attempts to describe the perturbative
degrees of freedom \cite{sbhh-polypert}). Most of the space-times studied to date in
LQC are either vacuum space-times \cite{mbgmmwe, gmbmm, mbmmwe} or with the particularly
simple choice of a massless scalar field \cite{aps-imp,apsv-spher, skl-spher, *v-open,%
*s-open, bp-negL, *pa-posL, awe-b1, awe-b2, *we-b9, *swe, mbmdbmm}.
While the cases of a massive scalar field in a flat FLRW
space-time \cite{aps-prep}, and a vector field in the Bianchi I cosmology \cite{al-vec}
have also been studied, a robust analysis of the dynamical sector of the theories 
at a genuinely quantum level has only been performed for matter choices (namely a
massless scalar field or pressure-less dust) which are idealizations of realistic (from the
point of view of particle physics) matter fields. In particular, the principal components
of the standard model have never been systematically analysed in this context and the 
LQC studies involving them rely on extrapolations from the above-mentioned idealizations.
In the general context of cosmology, a particularly important case is a perfect fluid of 
radiation formed by standard model particles.  This article is dedicated to the study
of a radiation-dominated space-time in LQC in the most simple setting possible: by emulating
a radiation-dominated perfect fluid with as few as possible homogeneous standard model matter 
fields.

As the resolution of cosmological singularities in LQC is due to the loop quantization
of the gravitational sector (rather than any effect due to the matter fields), it seems 
reasonable to assume that the specific type of matter field does not affect the qualitative 
results of LQC and that the big-bang singularity is generically replaced by a quantum gravity 
bounce, regardless of the matter field.
However, despite this expectation it is important to study a variety of matter
fields in order to show that the results obtained for massless scalar fields do 
in fact hold more generally, especially in the situation when distinct fields are used 
as the emergent time (i.e., the fields are used as evolution parameters labeling the 
families of partial observables \cite{obs-r,*obs-d,dkls-obs}), since the use of matter 
clocks has been essential in studies of the quantum dynamics to date%
\footnote{Once a particular matter field has been chosen as the internal 
  clock, it is easy to generalize the presence of the bounce and the energy
  density boundedness results to the case of generic matter fields
  \emph{added on top of} the clock fields. This is a direct consequence of
  the boundedness of the gravitational energy density operator 
  (see for example the discussion in \cite{hp-dust-LQC}).
}.

Another reason to study different matter fields is that it is not immediately obvious 
how to include some types of matter fields ---e.g., vector fields--- in minisuperspace 
models that assume homogeneity and isotropy.  This is a tricky problem as a homogeneous 
vector field necessarily picks out a preferred direction, which is clearly at odds 
with isotropy. We show that a model earlier proposed in classical cosmology 
is suitable for the Hamiltonian framework, and thus for canonical quantum cosmology 
theories such as LQC.

Finally, concerning the matter content, the massless scalar field mostly
used so far in the literature possesses a series of convenient properties that
simplify the analysis. As we will see later in this paper (Secs.~\ref{s.quant} 
and~\ref{s.asym} and Apps.~\ref{app:wdw} and~\ref{app:scatter}) other (more realistic) 
matter fields do not possess many of these properties. This makes the study of other 
fields significantly more difficult and in particular requires substantial
revisions and extensions of the analytical and numerical methods used in previous
studies, for example \cite{aps-imp}.  The increase in difficulty 
observed here for Maxwell fields is expected to be generic for 
standard model matter fields. Thus, the improvements to the methodology presented 
here may be essential for many further developments in LQC.

As a first step in addressing these issues, we will study the loop quantization of a
flat FLRW space-time with a Maxwell field, the massless vector field that satisfies
Maxwell's equations.  This is a particularly interesting matter field for two reasons.
First, it is a vector field, and therefore it will be necessary to determine how
a vector field can be handled in a homogeneous and isotropic background.  Second,
the equation of state of a Maxwell field is that of radiation, namely
$P = \rho/3$, where $\rho$ is the energy density and $P$ the pressure of
the Maxwell field.

This second condition is particularly important as all matter fields with the
dispersion relation $E = \sqrt{p^2 + m^2}$ that are in either the Bose-Einstein or
Fermi-Dirac distribution behave like radiation at sufficiently high temperatures,
that is they have the same equation of state $P \cong \rho/3$.  As the two
properties of the matter fields that enter into the Einstein equations in the
homogeneous and isotropic limit are precisely their energy density and pressure, different
fields that have the same initial $\rho$ and the same equation of state lead to the
same gravitational dynamics of the FLRW space-time.  Therefore, at high temperatures
(like the temperatures reached in the very early universe), all (of the populations) of the
bosonic or fermionic fields that can be treated in a statistical manner are
accurately mimicked by a population of Maxwell fields.  Thus, the study of the
Maxwell field in loop quantum cosmology is of wide interest, as it can provide a
good approximation to many different types of thermalized matter fields in the
Planck regime.

In this paper, following the improved dynamics loop quantization prescription
\cite{aps-imp}, we will study in detail the quantum dynamics of the isotropic 
universe with a suitable population of Maxwell fields as matter content and in 
particular show that the big-bang singularity is replaced by a quantum bounce
in LQC. This strongly suggests that the initial cosmological singularity is resolved
by quantum gravity effects in all flat FLRW space-times where the matter field
at high temperatures is well approximated by radiation.

The outline of the paper is as follows.  In Sec.~\ref{s.class}, the classical
theory will be reviewed and in particular it will be shown how, following
\cite{gmv-vec,*ap-vec}, a matter sector constituted of Maxwell fields can be
made to be homogeneous and isotropic.  Then in Sec.~\ref{s.quant} the
quantum theory will be defined and numerically solved for semi-classical
states; some results concerning the asymptotic dynamics are presented
in Sec.~\ref{s.asym}.  The effective equations are presented in Sec.~\ref{s.eff}, 
and we close with a discussion in Sec.~\ref{s.dis}. The technical derivations 
leading to the results presented in Sec.~\ref{s.asym} are contained in Apps.~%
\ref{app:wdw} and~\ref{app:scatter}.

The units we use in this paper are such that $c = 1$, but $G$ and $\hbar$
will remain explicit; we define the Planck length as $\lp = \sqrt{G \hbar}$.
Greek letters $\mu, \nu, \rho, \sigma$ represent space-time indices, while
the roman letters at the beginning of the alphabet $a, b, c$ represent
spatial indices and $i, j, k, l$ represent internal spatial indices.

\section{Vector Fields in Isotropic Cosmology}
\label{s.class}

In this section, we shall review a simple model that allows vector fields to be
included in homogeneous and isotropic minisuperspace models.  The key point is
that in a gas of photons, there are many individual photons evenly spread out
over space (ensuring approximate homogeneity) which are travelling in
all directions (ensuring approximate isotropy).  In the statistical limit of
a large number density of photons, the photon gas is homogeneous and isotropic.

One way to model the stress-energy tensor for a photon gas is by
working with a linear combination of plane waves.  In order to see this, recall that
the stress-energy tensor for a single plane wave of radiation with
amplitude $A$ and (null) tangent 4-vector $k^\mu$ is
\be
T_{\mu\nu} = \f{A^2}{8 \pi G} \, k_\mu k_\nu,
\ee
where we assume that we are interested in $T_{\mu\nu}$ only at scales larger than
the wavelength of the plane wave \cite{mtw}.

In order to satisfy the isotropy requirement, it is necessary to have plane waves
(with the same amplitude) travelling in all directions, and then
\be \label{int-plane-waves}
T_{\mu\nu} = \f{A^2}{8 \pi G} \, \int \dd \theta \dd \phi \sin \theta \,
k_\mu(\theta, \phi) k_\nu(\theta, \phi),
\ee
where $k^\mu(\theta, \phi)$ is the tangent 4-vector of the plane wave travelling
in the $(\theta, \phi)$ direction on the FLRW space-time with the metric
\be \label{eq:FLRW-metric}
\dd s^2 = -N^2 \dd t^2 + a(t)^2 \dd \vec{x}^2 .
\ee
From this, it is easy to check that
\be \label{eq:FLRW-Tab}
T_{\mu\nu} = \rho \, u_\mu u_\nu + P (g_{\mu\nu} + u_\mu u_\nu),
\ee
with
\be\label{eq:wave-pr}
\rho = \f{A^2}{2 G}, \qquad P = \f{\rho}{3},
\ee
and $u^\mu$ is the usual co-moving 4-vector of the perfect fluid.

This shows how it is possible to model a perfect fluid of radiation as
a linear combination of plane waves travelling in all directions.  However,
this setup is unwieldy in the Hamiltonian framework, so we will now introduce
a toy model that gives the stress-energy tensor \eqref{eq:FLRW-Tab}.  In
this toy model, there are three ``species'' or ``flavours'' of a Maxwell
field \cite{gmv-vec, ap-vec} in a flat FLRW universe.  This simpler setting
is relatively easy to handle in a Hamiltonian setting, and so is more
convenient for the quantum theory.  In the following part, we describe
this toy model, define the Hamiltonian for the matter and gravitational
sectors, and conclude the section by briefly discussing the classical
dynamics.

\subsection{The Three $U(1)$ Vector Fields}
\label{ss.c-matt}

Motivated by the fact that a linear superposition of plane waves can
yield a homogeneous and isotropic matter field, in this paper we will
consider a particularly simple linear superposition of this type.  To be
specific, we take a linear superposition of three plane waves that are
orthogonal and of equal amplitude.  Furthermore, for simplicity we assume
the wave number of these plane waves to be zero, which then each correspond
to {homogeneous field configurations}.  This
particularly simple linear superposition of plane waves is a toy model of
\eqref{int-plane-waves} which will make calculations
in a canonical quantum framework tractable and allow us to define the loop
quantum cosmology of a radiation-dominated space-time.  While a more realistic
model of a radiation-dominated space-time would be to consider a more general
linear superposition of plane waves, this is very difficult to handle in a
minisuperspace model of quantum cosmology and we leave this possibility for
future work. Nonetheless, \eqref{int-plane-waves} does suggest using a simpler
model ---namely, the linear superposition of the three {homogeneous} (i.e., plane
waves with zero momentum) and orthogonal vector fields--- in order to study a
radiation-dominated space-time%
\footnote{Note that this simpler model remains very interesting as it does
in fact capture the salient details of \eqref{int-plane-waves}.  This is
because the wave number does not affect the gravitational dynamics since
the energy density (as well as the pressure) contribution due to a plane
wave is homogeneous and only depends on the amplitude of the plane wave
as can be seen explicitly in \eqref{eq:wave-pr}.  Thus, the restriction
to one wave number will not affect the resulting physics insofar as the
dynamics of the space-time are concerned (where only energy density and
pressure enter into the Friedmann equations), and neither will the specific
choice of setting the wave number to zero in order to obtain homogeneous
solutions.}
This is what we shall do in this paper, and
in the remainder of this section we shall precisely define the model.

Denoting the vector potential of each field by $\aA_\mu$, where the index
$\al = 1, 2, 3$ labels the three different $U(1)$ fields, the Lagrangian density
for each of the three fields is given by
\be \label{lagrangian}
\aL = -\f{1}{4} \sqrt{-g} \, \aF_{\mu\nu} \aF^{\mu\nu},
\ee
where $\aF_{\mu\nu}$ is the field strength of the vector field,
$\aF_{\mu\nu} = 2 \partial_{[\mu} \aA_{\nu]}$.

Now, in order to ensure the homogeneity and isotropy of the matter field,
it is necessary to carefully choose the form of the $\aA_\mu$
for each $\al$.  To obtain a homogeneous and isotropic stress-energy
tensor, following \cite{gmv-vec,*ap-vec} we choose
\be \label{ansatz-u1}
\aA_a = \Ag(t) \de^\al_a, \qquad \aA_t = 0,
\ee
i.e., we take the three vector potentials to be mutually orthogonal, and
impose that they share the same time dependent length.
This choice gives a field strength where the only non-zero components
are
\be 
\aF_{ta} = - \aF_{at} = \partial_t \Ag \, \de^\al_a.
\ee

Given the metric \eqref{eq:FLRW-metric}, the Lagrangian density of the fields is
\be 
\aL = \f{a}{2N} (\partial_t \Ag)^2,
\ee
and then the canonical momentum of the vector field is given by
\be \label{momentum-u1}
{}^\al{\rm P} = \frac{\de (\aL)}{\de (\partial_t {}^\al A)}
= \f{a}{N} \aPi^{\mu},
\ee
where we have introduced
\be 
\aPi^{\mu} = (\partial_t \Ag) \delta^{\al\mu} \equiv \Pg \delta^{\al\mu}.
\ee
A subsequent Legendre transform gives the Hamiltonian density for one of the
$U(1)$ species,
\be 
\aH = \f{1}{2a} \Pg^2,
\ee
and the Poisson bracket between $\Ag$ and $\Pg$ is determined by the 
induced symplectic structure
\begin{equation} 
	\Omega(\de_1, \de_2) 
	= \sum_{\al = 1}^3 \int_{\Sigma} \Big( \de_1 \aA_\mu \de_2 \aPi^\mu
		- \de_2 \aA_\mu \de_1 \aPi^\mu \Big) \nn
	= 3 \Big( \de_1 \Ag \de_2 \Pg - \de_2 \Ag \de_1 \Pg \Big).
\end{equation}
The (appropriately regularized) integral over the spatial Cauchy slice $\Sigma$
can be performed trivially due to homogeneity, and the overall factor of the volume 
of the space can be absorbed into the definition of the fields%
\footnote{The integral has to be regularized as the space is non-compact. This is 
	done by introducing an infrared regulator, a compact co-moving spatial region 
	\cite{abl-lqc} (here a cubic cell $\mathcal{V}$ known as the \emph{fiducial cell}) 
	and subsequently ensuring that the resulting description admits a consistent 
	regulator removal limit \cite{aps-imp,awe-b1,cs-uniq,*cs-uniq2}. See also the 
	discussion in \cite{cm-reg,*cm-coh}.
}.
The Poisson bracket following from the induced symplectic structure is
\be \label{eq:poisson-matt}
\{ \Ag, \Pg \} = \f{1}{3}.
\ee

Finally, the total Hamiltonian density matter term is simply given by the sum of the
three individual Hamiltonian densities,
\be \label{eq:matter-ham}
\mH_m = \sum_{\al = 1}^3 \aH = \f{3}{2a} \Pg^2,
\ee
from which it is possible to calculate the energy density $\rho$ and the
pressure $P$ of the universe matter content:
\be \label{rho}
\rho = \f{\mH_m}{\sqrt{q}} = \f{3 \Pg^2}{2 a^4},
\ee
\be \label{pressure}
P = -\f{\partial \mH_m}{\partial {\rm Vol}}
= -\f{\partial \mH_m}{\partial a^3}
= \f{\Pg^2}{2 a^4}.
\ee
This implies in particular the relation $P = \rho/3$, just as one would expect 
for a radiation-dominated universe.

Alternatively, one can determine $P$ and $\rho$ by evaluating the stress energy tensor 
\cite{gmv-vec,*ap-vec}. That method also has the advantage of explicitly
showing that the stress energy tensor is that of a homogeneous and isotropic
perfect fluid.

\subsection{The Gravitational Sector}
\label{ss.c-grav}

Since we embed the matter fields discussed above in the isotropic flat spacetime,
the geometrical degrees of freedom are the scale factor $a$ and its canonical 
momentum $\ppi_{(a)}$ with the Poisson bracket $\{a,\ppi_{(a)}\}=1$. These variables
(together with the matter degrees of freedom) suffice for describing this symmetry-reduced
system (for details, see for example \cite{lr-frw}). However, there is another pair 
of conjugate variables that is more convenient for LQC (see Sec.~\ref{s.quant} and 
\cite{aps-imp,acs}) and which provide an equivalent description at the classical level, 
these variables are (proportional to) the oriented volume $\nu$ and its momentum $b$,
\begin{subequations}\label{eq:var-geom}\begin{align}
  \nu &= \f{a^3}{\al} ,  \hspace{3.2cm}
  \alpha = 2\pi\gamma \lp^2 \sqrt{\Delta} , \\
  b &= - \frac{2\alpha^{\frac{1}{3}}}{3\hbar} \cdot
    \frac{\ppi_{(a)}}{|\nu|^{2/3}} 
    = \frac{\alpha}{2\pi\lPl^2} H , \quad 
  \{\nu,b\} = -\frac{2}{\hbar} ,
\end{align}\end{subequations}
where the proportionality factor $\alpha$ contains the Barbero-Immirzi parameter 
$\gamma$ \cite{b-var} and the smallest non-zero eigenvalue of the LQG area operator 
$\Delta$.  $\Delta$ is often called the area gap and is of the order 
$\Delta \sim \lp^2$ \cite{al-area} (note that it has dimensions of area). The momentum 
$b$ is proportional to the Hubble parameter $H$ in the classical theory.

In the variables $(v,b)$ the gravitational term of the Hamiltonian constraint
density takes the form
\begin{equation}\label{eq:geom-ham}
  \mathcal{H}_g 
  = - \frac{\pi G}{3} \frac{\ppi_{(a)}^2}{a}
  = - \frac{3\pi G \hbar^2}{2 \alpha} |\nu|b^2 .
\end{equation}
Together with the matter Hamiltonian density \eqref{eq:matter-ham}, $\mH_g$
defines the classical dynamics of the system.

\subsection{Classical Dynamics}
\label{ss.c-dyn}

The classical dynamics of this model is generated by the Hamiltonian
constraint term in the canonical action
\begin{equation} \label{eq:preH}
N \mC_H = \int_\mV N \Big[ \mH_g + \mH_m \Big] , 
\end{equation}
where the gravitational and matter terms are given by \eqref{eq:geom-ham} and 
\eqref{eq:matter-ham} respectively. The physical trajectories lie on the surface
\begin{equation}
  N \mC_H = 0 .
\end{equation}

Note that the integration in \eqref{eq:preH} should in principle be performed over
the entire constant time slice $\Sigma$, however such an integral would diverge 
due to homogeneity and non-compactness of the slice. A standard way of removing 
this divergence is the introduction of an infrared regulator: a compact region
$\mathcal{V}$ of constant volume in comoving coordinates. Here for the
sake of simplicity we choose $\mathcal{V}$ to be the cubical cell of edges
generated by the vectors $\partial_x, \partial_y, \partial_z$ and of unit
volume with respect to the line element $\dd x^2 + \dd y^2 + \dd z^2$.
The equations of motion (presented below) do not
depend on the choice of the fiducial cell, and so it follows that
the physical results do not depend on the size of the cell and the limit of
removing the regulator is trivial in the classical theory. Note however
that this is not the case in the quantum theory where taking the limit of
$\mV \to \mathbb{R}^3$ is not trivial,
see \cite{cm-reg,rwe-eff} for more detailed discussions on this point.
Nonetheless, this limit exists and the resulting quantum theory
is independent of the initial choice of the fiducial cell.

Performing the (regulated) integral in \eqref{eq:preH} we arrive at the following
form of the constraint,
\be 
  \mC_H = -\f{3\pi G\hbar^2}{2\alpha}|\nu| b^2
  + \f{3}{2 |\al \nu|^{1/3}} \Pg^2 .
\ee
So far the choice of the lapse $N$ remains open, however in the quantum theory we
will want to deparametrize the system with respect to the matter field in order to
use $\Ag$ as a clock. To synchronize the classical time with that clock, we choose 
$N = a(t) = \al |\nu|^{1/3}$ and denote the resulting time variable $\eta$. This
choice corresponds to working in conformal time, and the resulting constraint 
reads
\be \label{c-ham}
  N \mC_H = -\f{3\pi G\hbar^2}{2\alpha^{2/3}}|\nu|^{4/3} \, b^2 
  + \f{3}{2} \Pg^2.
\ee
The equations of motion are the Hamilton-Jacobi equations which in this case consist
of the following set of four coupled ordinary differential equations,
\begin{subequations}\label{eq:eom}\begin{align}
  \f{\rd v}{\rd\eta} 
    &= \f{6\pi G\hbar}{\alpha^{2/3}} |\nu|^{4/3} \, b ,  &
  \f{\rd b}{\rd\eta} 
    &= - \f{4 \pi G\hbar}{\alpha^{2/3}} \sgn(\nu) |\nu|^{1/3} \, b^2 ,
    \label{eq:eom-geom} \\
  \f{\rd A_{\gamma}}{\rd\eta} &= \Pi_{\gamma} ,  &
  \f{\rd \Pi_{\gamma}}{\rd\eta} &= 0 . 
    \label{eq:eom-matt}
\end{align}\end{subequations}
These equations are equivalent to the usual Friedmann equations and are easily
solved.

The matter degrees of freedom are determined by \eqref{eq:eom-matt}: the
momentum of the electromagnetic field is a constant of the motion and $\Ag$ grows
linearly in conformal time
\be \label{c-sol-matt}
\Pg = {\rm const}, \qquad \Ag(\eta) = \Pg \eta + \Pi_o,
\ee
where $\Pi_o$ is a free constant of integration.

The dynamics of $(v,b)$ are determined by \eqref{eq:eom-geom} and can be found 
in two steps: first we note that the equation for an auxiliary variable 
$f:=b|v|^{1/3}$ decouples from the system and so can easily be solved. 
Then, the $(v,b)$ variables can be found once the solution $f$ is plugged
back into \eqref{eq:eom-geom} with the result
\begin{subequations}\label{c-sol-grav}\begin{align}
  \nu(\eta) &= \f{(4 \pi G)^{3/2}}{\al} \, \Pg^3 \, (\eta - \eta_o)^3 , & 
  b(\eta) &= \frac{\al}{2 \pi G \hbar \sqrt{4 \pi G}} \cdot
     \frac{1}{\Pg \, (\eta-\eta_o)^2},
\end{align}\end{subequations}
where $\eta_o$ is a constant of integration corresponding to the
moment of initial/final singularity%
\footnote{The solutions (\ref{c-sol-matt}, \ref{c-sol-grav}) contain two branches:
  $(\eta>\eta_o)$ representing the expanding universe starting at the initial 
  singularity $\eta=\eta_o$ and $(\eta<\eta_o)$ representing the universe 
  contracting to the final singularity at $\eta=\eta_o$. Due to the irregularity of
  the equations of motion at $\eta=\eta_o$ there is no unique extension of the
  solution across the point $\eta=\eta_o$, although there is a unique analytic
  extension through that point.
}.
Note that there is another constant of integration that is fixed by enforcing
the constraint $N \mC_H = 0$.

The monotonicity of $\Ag(\eta)$ allows us to elimitate the time dependence 
from the equations of motion by reparametrizing the evolution in terms of $\Ag$
and thus use the vector potential as an internal clock,
\begin{subequations} \label{eq:class-traj-dep} \begin{align}
  \nu(\Ag) &= \f{(4 \pi G)^{3/2}}{\al} \, \left( \Ag - A_o \right)^3, &
  b(\Ag) &= \frac{\al}{2 \pi G \hbar \sqrt{4 \pi G}} \cdot
      \frac{\Pg}{(\Ag - A_o)^2},
\end{align} \end{subequations}
where $A_o = \Pi_o + \Pg \eta_o$.  $A_o$ then represents the ``initial time''
where the big-bang or big-crunch singularity occurs, according to the $\Ag$
clock.

Finally, the energy density and pressure can be determined from
Eqs.~\eqref{rho} and \eqref{pressure}:
\be 
\rho = \f{3}{32 \pi^2 G^2 \Pg^2 (\eta - \eta_o)^4}
= \f{3 \, \Pg^2}{32 \pi^2 G^2 (\Ag - A_o)^4},
\ee
\be 
P = \f{1}{32 \pi^2 G^2 \Pg^2 (\eta - \eta_o)^4}
= \f{\Pg^2}{32 \pi^2 G^2 (\Ag - A_o)^4} = \f{\rho}{3}.
\ee

\section{The Quantum Theory}
\label{s.quant}

The procedure of quantizing (within the LQC framework) the classical system specified 
in the previous section is a direct analog of the procedure used for isotropic
systems with a scalar field \cite{aps-imp,awe-b1}. Therefore we recall it here only briefly,
focusing mainly on the differences with respect to previous treatments and on the
specific steps where quantization ambiguities force us to make particular choices.

\subsection{The Dirac Program}

In general the process is an application of the Dirac program: first the constrained 
system is quantized while ignoring the constraints (the so-called \emph{kinematical} 
quantization), then the constraints are defined as quantum operators, and finally 
the space of physical states is constructed out of the kernel of the quantum constraint
operators. In this setting, meaningful physical quantities are represented
by \emph{partial observables} \cite{obs-r,*obs-d}.
Let us start by recalling the kinematical quantization.

\subsubsection{The Kinematical Hilbert Space}
\label{sss.hilbert}

In this step, following the majority of works in LQC, we implement a hybrid
approach, quantizing the geometry degrees of freedom via a polymeric
quantization while using the standard quantum mechanical tools for the matter
sector \cite{aps-prl}. Thus, the kinematical Hilbert space is a product
$\Hil_{\rm kin} = \Hil_{\rm gr} \otimes \Hil_{A}$ of the gravitational and
matter Hilbert spaces.

The gravitational Hilbert space is the space of square-summable functions on the 
Bohr compactification of the real line (the space of 
almost periodic functions) $\Hil_{\rm gr} = L^2(\bar{\re},\rd\mu_{\rm Bohr})$.
A convenient basis for this Hilbert space is formed by the eigenfunctions of
the $\wh{v}$ operator%
\footnote{Note that from the $\wh{v}$ operator, it is possible to
		construct the operator $\wh{p} = \sgn(v)|\alpha \wh{v}|^{2/3}$,
		which corresponds to the flux of the densitized triad across one
		of the faces of the fiducial cell.},
the quantum counterpart of the variable $v$ introduced in \eqref{eq:var-geom}.
The inner product on $\Hil_{\rm gr}$ is discrete,
\begin{equation}
  \langle v | v' \rangle = \delta_{v,v'} ,
\end{equation}
where however $v$ runs through the whole real line. As a consequence, $\Hil_{\rm gr}$
is nonseparable.

Normalizable states on $\Hil_{\rm gr}$ are represented by the wave function $\psi(v)$,
\begin{equation}
  |\psi\rangle = \sum_{v\in\re} \psi(v)|v\rangle , \qquad \sum_{v\in\re} |\psi(v)| <\infty .
\end{equation}
We will require that operators acting within this space be well-defined on the domain
$\Dom_{\rm gr}$ of \emph{finite} linear combinations of $|v\rangle$. 

As the basic operators defined on a dense domain in $\Hil_{\rm gr}$ we choose the
operator $\wh{v}$ (proportional to the volume of the chosen comoving region of space)
and the $U(1)$ unit shift operator $\wh{N}$ such that
\begin{equation}\label{eq:op-gr}
  \wh{v}|v\rangle = v|v\rangle , \qquad \wh{N}|v\rangle = | v+1\rangle.
\end{equation}
The standard elementary operators of isotropic LQC ---namely, the triad flux $\wh{p}$ 
across a face of the fiducial cell and the $SU(2)$ holonomy $h_{\lambda}$ along a straight 
line of fiducial length $\lambda$--- can be expressed in terms of $\wh{v},\wh{N}$
\cite{aps-imp,acs}.

The matter Hilbert space is the standard Lebesgue space $\Hil_{A} = L^2(\re,\rd\Ag)$. 
As a basis we choose the generalized eigenstates $(\Ag|$ of the field operator $\wAg$.
States are represented by square-integrable wave functions
$\psi(\Ag) := (\Ag|\psi\rangle$ and
the basic operators are
\begin{subequations} \label{eq:op-matt} \begin{align}
\wAg \psi(\Ag) &= \Ag \psi(\Ag), &
\wPg \psi(\Ag) &= -\f{i\hbar}{3} \f{\dd}{\dd \Ag} \psi(\Ag); 
\end{align}\end{subequations}
the domain on which we require the operators to be well-defined is the 
Schwartz space $\mathcal{S}(\re)$.

\subsubsection{The Hamiltonian Constraint Operator}
\label{sss.q-ham}

The next step in the Dirac program is the construction of the quantum operator
representing the Hamiltonian constraint \eqref{c-ham} and composed of the basic 
kinematical operators defined in \eqref{eq:op-gr} and \eqref{eq:op-matt}.
The procedure, while a bit complicated, is well descibed in the literature, see 
for example \cite{aps-imp,awe-b1}.
To capture the properties of full LQG we start by expressing the constraint 
\eqref{c-ham} in terms of the $SU(2)$ holonomies $A^i_a$ and densitized triads 
$E^a_i$ directly. Next the constraint is regularized following the prescription
given by Thiemann \cite{thiemann}.  In particular, the field strength (i.e.,
the curvature of the connection) is expressed in terms of a
holonomy along a closed square loop of physical area equal to the 
lowest non-zero eigenvalue of the area operator in LQG (relevant for LQC)
$\Delta = 4\sqrt{3}\pi\gamma \lp^2$.
As a result the gravitational part of the constraint is expressed in terms of the
holonomy functions and the volume, which next are promoted to composite operators
expressed in terms of the operators \eqref{eq:op-gr}. The matter part of the
constraint does not need any special treatment and can be immediately expressed in
terms of the operators \eqref{eq:op-matt}.

The last step listed here involves some ambiguity due to different
factor-ordering choices that are possible.
Here we choose a particularly convenient factor-ordering motivated by
studies of the anisotropic Bianchi I cosmology
\cite{awe-b1} which involves a specific treatment of the sign function and
simplifies the resulting physical Hilbert space structure \cite{mmo-FLRW}. The
final form of the operator is%
\footnote{It is defined analogously to the prescription 
	provided for the system with a massless scalar field in \cite{mop-presc} and denoted 
	there as \emph{sMMO}. See \cite{mop-presc} for its description and a comparison with 
	other factor-ordering choices.}
\begin{equation}\label{eq:KG-LQC}
  \widehat{N\mathcal{C}_H} = \Theta \otimes \Id 
  + \Id \otimes \frac{\partial^2}{\partial\Ag^2} ,
\end{equation}
where the operator $\Theta$ (also called the evolution operator) 
takes the form
\begin{widetext}
\begin{equation}\label{q-cg-action}\begin{split}
\Theta \Psi(\nu; \Ag) =
- \f{9 (2 \pi \ga \sqrt\Delta)^{1/3}}{32 \ga \sqrt\Delta \hbar} |\nu|^{1/3}
\times & \Bigg[ s_-(\nu-2) s_-(\nu) |\nu-4|^{1/3} |\nu-2|^{2/3} \Psi(\nu-4; \Ag)  \\
& - \bigg( s_-^2(\nu) |\nu-2|^{2/3} + s_+^2(\nu) |\nu+2|^{2/3} \bigg)
|\nu|^{1/3} \Psi(\nu; \Ag)   \\
& + s_+(\nu+2) s_+(\nu) |\nu+4|^{1/3} |\nu+2|^{2/3} \Psi(\nu+4; \Ag) \Bigg],
\end{split}\end{equation}
\end{widetext}
where $s_{\pm}(\nu) = \sgn(\nu \pm 2) + \sgn(\nu)$. This particular form
of $\Theta$ has several convenient properties:
\begin{enumerate}
  \item
    The zero volume state $\ket{\nu = 0}$ decouples under the action of
    $\h\mC_H$. 
  \item
    Due to presence of the $s_\pm(\nu)$ terms the sectors $\nu > 0$ and $\nu < 0$
    decouple. 
  \item
    Since $\Theta$ is a difference operator coupling only points in $\nu$ separated 
    by $4$, one can split the support of $\Hil_{\rm gr}$ elements onto independent 
    sets $\nu = \epsilon + 4n$ (preserved under the action of $\h\mC_H$),
    where $n \in \mathbb{Z}$ and $\epsilon \in (0, 4]$.
\end{enumerate}
The first property implies that we can exclude the singular $\ket{v=0}$ states
from the support of the wave function, showing that the singularity is resolved
at the quasi-kinematical level.  

At this point it is useful to note one more important property of the model, 
namely that it does not feature parity violating interactions. In consequence the 
triad reflection (here represented by $\nu\mapsto -\nu$) is a large symmetry and
the subspaces of symmetric and antisymmetric (with respect to that reflection) 
states are superselection sectors. As a consequence one can choose just one of
them and then, due to properties $1$ and $2$ above, restrict the support of the
wave function to $\nu>0$.

That restriction, together with property $3$, allows us to divide
$\Hil_{\rm gr}$ into superselection sectors consisting of the 
projections of $\psi\in\Hil_{\rm gr}$ onto the positive semi-lattices
\begin{equation}
  \mathcal{L}_{\epsilon} = \{ v\in\re: v = \epsilon +4n, n\in\mathbb{N} \},
\end{equation}
and work with just a single superselection sector, provided that these sectors are 
also preserved by the chosen set of observables (which as we shall see below is 
the case). Of course, it is important to verify that the physics does not depend 
on the choice of the superselection sector, as there does not exist any principle 
that could justify one choice over another.

One justification for working with a single sector is that while the entire Hilbert 
space $\Hil_{\rm gr}$ is not separable, each superselection sector
$\Hil_{{\rm gr},\epsilon}:=\{\psi|_{\mathcal{L}_{\epsilon}}; \psi\in\Hil_{\rm gr} \}$
is. An alternative possibility to construct a separable Hilbert space is
to use the construction given in Appendix~C of \cite{aps-det} or to exploit the natural 
fibration of $\Hil_{\rm gr}$ and the Lebesgue measure on the fiber space inherited 
from superselection labels. The latter method leads to the fiber-integral Hilbert space
which is again separable \cite{kp-polysc,bpv-osc}. For the remaining part of this
paper we choose the first approach and work with one superselection sector.

Upon restriction to a positive semi-lattice $\mathcal{L}_{\epsilon}$, the evolution operator
$\Theta$ [whose action is given in \eqref{q-cg-action}] can be simplified to
\begin{widetext}
\begin{equation} \label{Theta-op}\begin{split}
	\Theta \Psi(\nu; \Ag) 
	&= f^+(\nu) \Psi(\nu+4; \Ag) + f^o(\nu) \Psi(\nu; \Ag) f^-(\nu) \Psi(\nu-4; \Ag) \\
	&= - \f{9 (2 \pi \ga \sqrt\Delta)^{1/3}}{8 \ga \sqrt\Delta \hbar} \nu^{1/3}
     \times \Bigg[ \theta(\nu-4) (\nu-4)^{1/3} (\nu-2)^{2/3} \Psi(\nu-4; \Ag)  \\
	& \hspace{4.7cm}  - \bigg( \theta(\nu-2) (\nu-2)^{2/3} + (\nu+2)^{2/3} \bigg)
    \nu^{1/3} \Psi(\nu; \Ag)  \\
  & \hspace{4.7cm} + (\nu+4)^{1/3} (\nu+2)^{2/3} \Psi(\nu+4; \Ag) \Bigg],
\end{split}\end{equation}
\end{widetext}
where $\theta(\nu)$ is the Heaviside step function.

From this one can infer several important properties:
\begin{enumerate}[a)]
	\item Since $f^\pm,f^o$ are real functions, the operator is real.
	\item All (generalized) eigenfunctions are solutions to the second
		order difference equation
		\begin{equation}\label{eq:eig-gen}
			\Theta \psi_{\lambda} (\nu) = \lambda \: \psi_{\lambda} (\nu) .
		\end{equation}
	\item Due to the presence of $\theta$ functions in \eqref{Theta-op},
	  $\psi_{\lambda}(\epsilon+4)$ is uniquely determined by $\psi_{\lambda}(\epsilon)$,
	  therefore all the eigenspaces are $1$-dimensional and the spectrum ${\rm Sp}(\Theta)$
	  is nondegenerate.
	\item Due to the reality of the operator, all of its eigenfunctions are real up to a 
		global phase.
	\item By construction, the operator is symmetric and positive definite on its 
		domain $\Dom$.
	\item By direct inspection of its deficiency functions (using numerical methods 
		described in \cite{aps-imp,mop-presc} and the asymptotic analysis of \cite{kp-scatter}, 
		further applied in Sec.~\ref{sss.lqc-scatt}), one can show that $\Theta$ is essentially 
		self-adjoint.
	\item By exploring the asymptotic properties of the generalized eigenfunctions 
		corresponding to the positive eigenvalues $\lambda=\omega^2$ and the spectral 
		properties of the WDW analog of $\Theta$, one can show that%
			\footnote{Unfortunately, as the 
			leading power of $v$ in $f$ for large $\nu$ is not an integer,
			the mathematically precise proof of this property presented in \cite{kl-flat-sadj}
			for a massless scalar field cannot be adapted to the case of a Maxwell
			field that is considered here.}
		the spectrum of $\Theta$ is purely continuous and ${\rm Sp}(\Theta)=\re^+$.

\end{enumerate}

These properties will be essential in constructing the physical Hilbert space and probing the dynamics of the system.

\subsubsection{The Physical Hilbert Space}
\label{sss:hil-phys}

In the Dirac program, the space of physical states is defined as the set of states annihilated 
by the quantum constraint operator, that is $\widehat{N\mC_H}\Psi=0$. Thus,
these states must satisfy the equation
\be \label{q-time}
- \f{\partial^2}{\partial \Ag^2} \Psi(\nu; \Ag) = \Theta \Psi(\nu; \Ag),
\ee
where the action of $\Theta$ is given in \eqref{Theta-op}.

A systematic way to find this space is the so-called group averaging
method \cite{almmt-gave,*m-gave1,*m-gave2,*m-gave3,*m-gave4}, where
one builds an antilinear rigging map that provides an ``extractor operator''
$\mathcal{P}$ which projects the kinematical state $\Phi$ onto the physical
wave function $\Psi$:
\begin{equation}
  \Psi(\nu,\Ag) = [\mathcal{P} \Phi] (\nu,\Ag) = \int \rd t \: e^{it\widehat{N\mC_H}} \:
  \Phi(\nu,\Ag) .
\end{equation}
Using the spectral properties of $\Theta$ and $\partial_{\Ag}^2$ and following
the algorithm specified in \cite{aps-det}, one can easily determine
the form of the physical states, 
\begin{equation}\label{eq:gave-phy}
	\Psi(\nu,\Ag) = \int_{\re^+} \rd k \: \tilde{\Psi}^+(k) \: e_k(\nu) \: e^{i\omega(k)\Ag} \\
	  + \int_{\re^+} \rd k \: \tilde{\Psi}^-(k) \: \bar{e}_k(\nu) \: e^{-i\omega(k)\Ag} ,
\end{equation}
where $\tilde{\Psi}^\pm(k)\in L^2(\re^+,\rd k)$ are the spectral profiles of what
we shall call the positive and negative ``frequency'' components%
\footnote{The reason for choosing this nomenclature will become 
	clear in the next section.},
the norm is $\|\Psi\|^2 = \|\tilde{\Psi}^+\|_{L^2}^2 + \|\tilde{\Psi}^-\|_{L^2}^2$
and $e_k$ are the Dirac delta normalized (generalized) eigenfunctions of $\Theta$
that can be chosen to be real and with $e_k(\epsilon)>0$,
\begin{equation}\label{eq:eig}
  \Theta \: e_k = \omega^2(k) \: e_k,
\end{equation}
with the relation between the ``wave'' label $k$ and the ``frequency'' $\omega$
being given by%
\footnote{At the moment this relation is arbitrary as we can relabel $k$,
	although this particular choice is justified by the asymptotic properties of 
	$e_k$ analyzed in App.~\ref{sss.lqc-scatt}.}
\be \label{lqc-eigvalue}
	\omega^2_k = 2 \f{(2 \pi \ga \sqrt\Delta)^{1/3}}
	{\ga \sqrt\Delta \hbar} k^2.
\ee
Using the similarity of \eqref{q-time} to the Klein-Gordon equation (with $\Ag$ being 
the analog of time) one can conclude that the first and second components of 
\eqref{eq:gave-phy} are the ``frequency'' superselection sectors, of which we
select the first one.  In this case the equation \eqref{q-time} can be
rewritten in the form
\begin{equation} \label{eq:lqc-schroed}
- i \f{\partial}{\partial \Ag} \Psi(\nu; \Ag)
= \sqrt{\Theta} \, \Psi(\nu; \Ag).
\end{equation}
Finally, the physical states are
\begin{equation}\label{eq:psi}
  \Psi(\nu,\Ag) = \int_{\re^+}\rd k \: \Psi(k) \: e_k(\nu) \: e^{i\omega(k)\Ag} ,
\end{equation}
and the physical inner product is
\begin{equation}
	\langle\Psi|\Psi'\rangle = \int_{\re^{+}}\rd k \: \overline{\Psi}(k) \: \Psi'(k).
\end{equation}

The problem with this construction is the standard one that arises in
constrained systems: this gives a frozen time evolution where physical states
represent entire ``histories'' of the universe.  Providing a physically 
meaningful and non-trivial notion of evolution is not straightforward.

\subsection{Physical Evolution}

There are two well-developed methods that can be used in order to define meaningful
dynamics in a constrained system, namely \emph{deparametrization} and the \emph{partial 
observable} formalism. For the system considered here these two methods are equivalent,
although this equivalence does not hold in general. In fact, even in isotropic
LQC there exist systems where these two approaches give distinct results \cite{klp-ga}.

\subsubsection{Deparametrization}

The principal idea behind deparametrization is the observation that all the information 
about the physical state $|\Psi\rangle$ is contained within a single constant $\Ag$
slice of the wave function. Indeed, the Schr\"odinger-like equation \eqref{eq:lqc-schroed}
implies that given an initial slice $\Psi(\cdot,\Ag)\in\Hil_{\rm gr}$, we can reproduce
the entire physical state via the unitary transformations
\begin{equation}\label{eq:uni}
  \Psi(\nu,\Ag) = U_{\Ag{}_o,\Ag} \Psi(\nu,\Ag{}_o) 
  = e^{i\sqrt{\Theta}(\Ag-\Ag{}_o)} \Psi(\nu,\Ag{}_o) .
\end{equation}
The structure of the physical states \eqref{eq:psi} and the properties of $\Theta$ 
(namely its self-adjointness and positive definiteness) also imply the equivalence 
between the physical inner product and the gravitational kinematical inner product
on constant $\Ag$ slices
\begin{equation}\label{eq:ip-equiv}
  \forall \Ag\in\re:\ 
  \langle\Psi|\Psi'\rangle_{\rm phy} 
  = \langle\Psi(\cdot,\Ag)|\Psi'(\cdot,\Ag)\rangle_{\rm gr} ,
\end{equation}
which allows us to interpret $\Ag$ as an \emph{emergent time} and use it as an 
evolution parameter. We can thus define the dynamics as the unitary mapping
\begin{equation}\label{eq:evo-map}
  \re \ni \Ag \mapsto \Psi(\cdot,\Ag)\in\Hil_{\rm gr} .
\end{equation}
A convenient consequence of this approach is the fact that any self-adjoint operator
on $\Hil_{\rm gr}$ evaluated at a particular ``time'' $\Ag$ automatically
becomes a physical observable.

This property allows us to easily show that there exists an upper bound on 
the physical matter energy density. To show this, we start by defining
in the kinematical Hilbert space the operator corresponding to the matter energy
density defined in \eqref{rho},
\be \label{rho-op}
	\wh \rho = \f{3}{2} \left[ \Big( \wh{|\alpha \nu|} \Big)^{-1/3} \wPg
	\Big( \wh{|\alpha \nu|} \Big)^{-1/3} \right]^2.
\ee
As the zero-volume states have decoupled, the inverse $|\alpha \nu|$ operator
is well-defined.  This form of the $\wh \rho$ operator clearly shows that
it is a positive-definite operator as it is the square of a self-adjoint
operator.  Therefore, all of its eigenvalues will be real and greater than
or equal to zero.

In addition, for states annihilated by the constraint \eqref{q-time},
the action of this operator is exactly balanced via \eqref{eq:lqc-schroed}
by the gravitational energy operator which (after selecting a convenient factor
ordering) can be writen as \cite{acs,klp-aspects,hp-dust-LQC}
\begin{equation}\label{eq:rho-g}
  \rho_G = -\tilde{\rho} \otimes \Id , \qquad \tilde{\rho} = \rho_c \sin^2(b),
\end{equation}
where the critical energy density equals
\be \label{eq:rhoc}
  \rho_c = \frac{3}{8 \pi \ga^2 \Delta G},
\ee
and is of the order of the Planck energy density%
\footnote{In the numerical studies we took the value determined
    via \eqref{eq:rhoc} for the value $\gamma=0.2375...$ 
	determined in \cite{dl-gamma} ,which gives
    $\rho_c\approx 0.41\rho_{{\rm Pl}}$.}.

Since the operator $\tilde{\rho}$ is self-adjoint on $\Hil_{\rm gr}$, in the 
deparametrization picture it becomes a physical observable. On the other hand,
its form implies immediately that
\begin{equation}
  {\rm Sp}(\tilde{\rho}) = [0,\rho_c] .
\end{equation}
Therefore, the physical matter energy density is bounded from above by $\rho_c$.

Note that in order to determine whether this upper bound is saturated one needs 
to study in detail the dynamical evolution of the state in the sense of
\eqref{eq:evo-map}.

\subsubsection{The Partial Observable Formalim}

An alternative way to define the dynamics is provided by the formalism of partial 
observables originally introduced in \cite{obs-r,*obs-d} (see also \cite{dkls-obs} 
for a critical analysis of the formalism). The goal is to construct Dirac observables 
acting within the physical Hilbert space, where these observables correspond
to measuring one property of the state with respect to another. Here the natural
choice is to measure quantities with respect to the vector potential $\Ag$,
which will act as a relational clock.

These observables can be constructed systematically out of observables in the kinematical
Hilbert space via group averaging \cite{klp-ga}. In particular, in our case, given a
self-adjoint operator $\wh{O}$ on $\Hil_{\rm gr}$, the group averaging of the ``seed''
operator $\wh{O}\otimes\delta(\Ag-T)$ gives the Dirac observable $\wh{O}_{T}$
\begin{equation}
  \wh{O}_T \Psi(\nu,\Ag) = e^{i\sqrt{\Theta}(\Ag-T)} \wh{O} \Psi(\nu,T) , 
\end{equation}
which is just a completion of the (result of the) action of the gravitational 
kinematical observable $\wh{O}$ to the physical state \eqref{eq:psi} via the unitary 
transformation \eqref{eq:uni}. In consequence, in this setting the formalism
is equivalent to the deparametrization picture.

Using this procedure we construct the following set of observables: the volume 
$\wh V$ at the ``time'' $\Ag|_0$ and the momentum $\wPg$ (which is a constant
of the motion), given by
\begin{subequations}\begin{align}
		\wh{V(\Ag|_0)} \Psi(\nu; \Ag) &= e^{i \sqrt{|\Theta|} (\Ag - \Ag|_o)}
		(2 \pi \ga \sqrt\Delta \lp^3) 
		\, \nu \Psi(\nu; \Ag|_0), 
	\\
	\wPg \Psi(\nu; \Ag) &= -i\hbar \f{\partial}{\partial \Ag} \Psi(\nu; \Ag).
\end{align}\end{subequations}

\subsubsection{Dynamics}
\label{ss.num}

Either one of the two formalisms discussed in the previous sub-subsections can be 
used to determine the dynamics of this system. To achieve this goal we need to perform 
two tasks: the evaluation of the wave functions and the computation of the expectation 
values and the dispersions of the relevant observables.

To determine the wave function $\Psi(\nu,\Ag)$ as given in \eqref{eq:psi} one 
needs to know the explicit form of $e_k$. Unfortunately the form of $\Theta$
is not simple enough to easily 
determine its analytic form. We therefore resort to numerical methods, applying 
directly the techniques specified in \cite{aps-det,mop-presc}: solving directly
\eqref{eq:eig}, which after substituting the form \eqref{Theta-op} of $\Theta$ 
becomes a second%
\footnote{Its space of solutions is nonetheless
	$1$-dimensional due to the orientation decoupling
	(see the discussion in Sec.~\ref{sss.q-ham}).
}
order difference equation, and normalizing the solution via its WDW limit analysis 
(see the discussion in App.~\ref{app:scatter}). The normalization procedure features 
the only difference with respect to the treatment of \cite{aps-det,mop-presc}.
Namely, due to slightly more complicated structure of the WDW analog of our model
and a lower order of convergence of $e_k$ to its WDW component limit, we used the 
modified auxiliary basis functions \eqref{eq:lim-imp} instead.

The wave function itself is then determined via a direct numerical integration 
(using the Romberg method) of \eqref{eq:psi} (see again \cite{aps-det,mop-presc}
for a detailed description of the method).

In the actual computations we used the Gaussian spectral profiles 
\begin{equation}\label{eq:G-prof}
  \tilde{\Psi}(k) = N e^{-(k-k_o)^2/2\sigma^2} e^{ik\Ag^\star}
\end{equation}
peaked about $k_o$ and with $\Ag^\star$ selected to reproduce the value 
of expectation value of observable $\widehat{V(\Ag)} = V^\star$ for the chosen
$V^\star$ at the chosen ``initial time'' $\Ag = \Ag^o$. In practice we use the 
formula following from the classical trajectory \eqref{eq:class-traj-dep} which 
reproduces the desired result sufficiently well so long as the parameters 
satisfy $V^\star \gg |k^\star|\lPl^3$.

For the chosen profile \eqref{eq:G-prof} the expectation values and dispersion of 
$\wPg$ can be determined analytically,
\begin{equation}
  \langle \wPg \rangle = |k_o| , \qquad
  \langle \Delta \wPg \rangle = \sigma .
\end{equation}
To evaluate the analogous quantities for the observables $\widehat{V(\Ag)}$ we 
use the deparametrization picture, evaluating 
$\langle \Psi(\cdot,\Ag) |\wh{V} | \Psi(\cdot,\Ag) \rangle_{\rm gr}$ and
$\langle \Psi(\cdot,\Ag) |\Delta\wh{V} | \Psi(\cdot,\Ag) \rangle_{\rm gr}$ 
by direct numerical summation (see again \cite{aps-det,mop-presc}). 

In the studies performed the parameters were chosen from within the ranges
$\Pg\in [30,1700] \, (G\sqrt{\hbar})^{-1}$, $\Delta\Pg/\Pg\in [0.02,0.1]$, 
$V(\Ag^o) \in [3.1\cdot 10^3, 2.5\cdot 10^4] \, \lPl^3$.

\begin{figure}[tbh!]
	\includegraphics[width=0.96\textwidth]{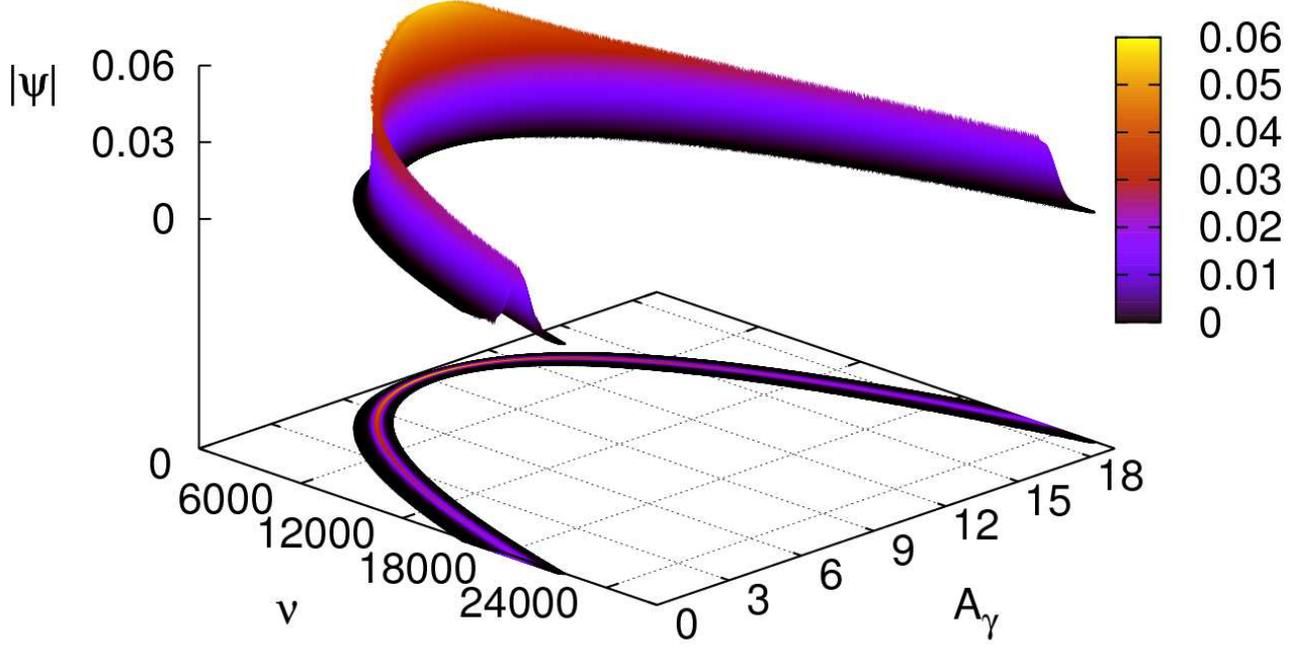}
	\caption{An example of a physical state: forward evolution of the Gaussian 
	\eqref{eq:G-prof} packet peaked about $\Pg\approx 83.3 \, (G\sqrt\hbar)^{-1}$
           and (at initial time $\Ag=0$) about $\nu\approx 1.8\times10^4$.}
    \label{fig:dynamics-3d}
\end{figure}

\begin{figure*}[tbh!]
	\subfigure[]{\includegraphics[width=0.49\textwidth]{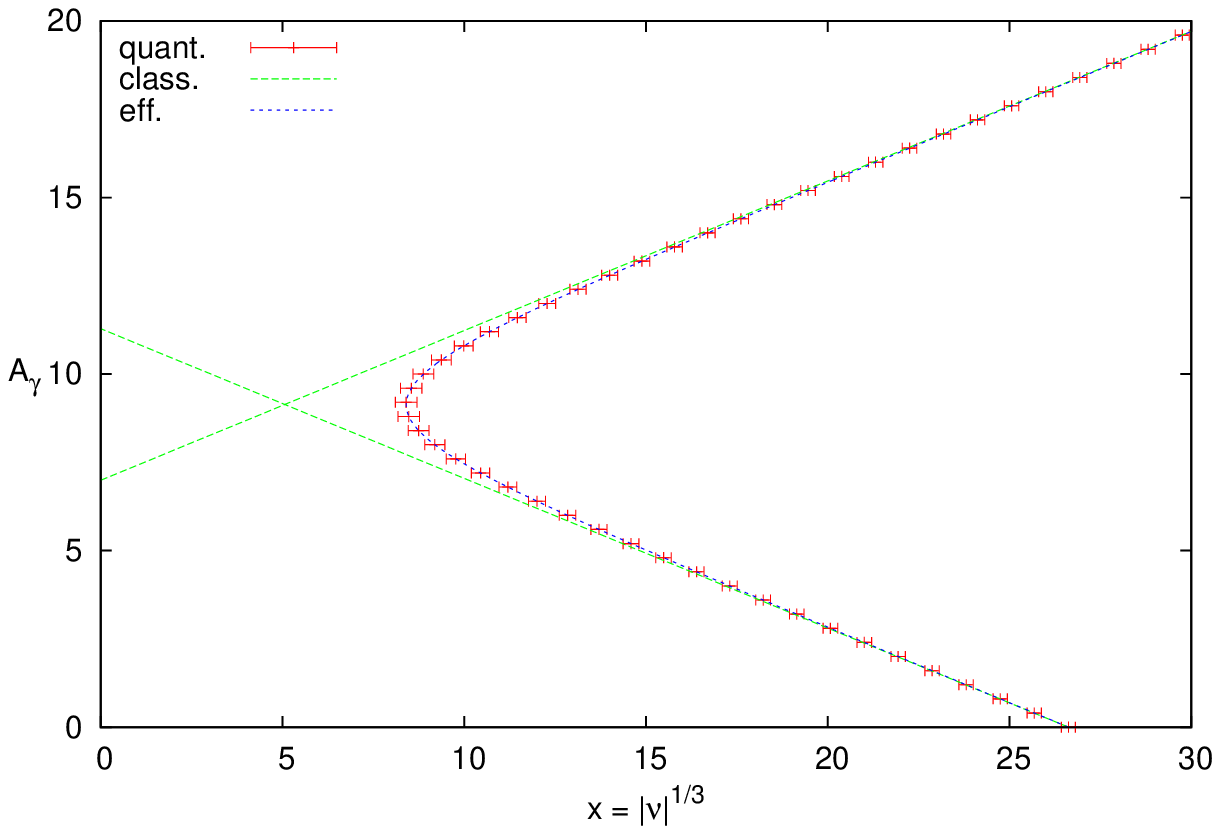}}
	\subfigure[]{\includegraphics[width=0.49\textwidth]{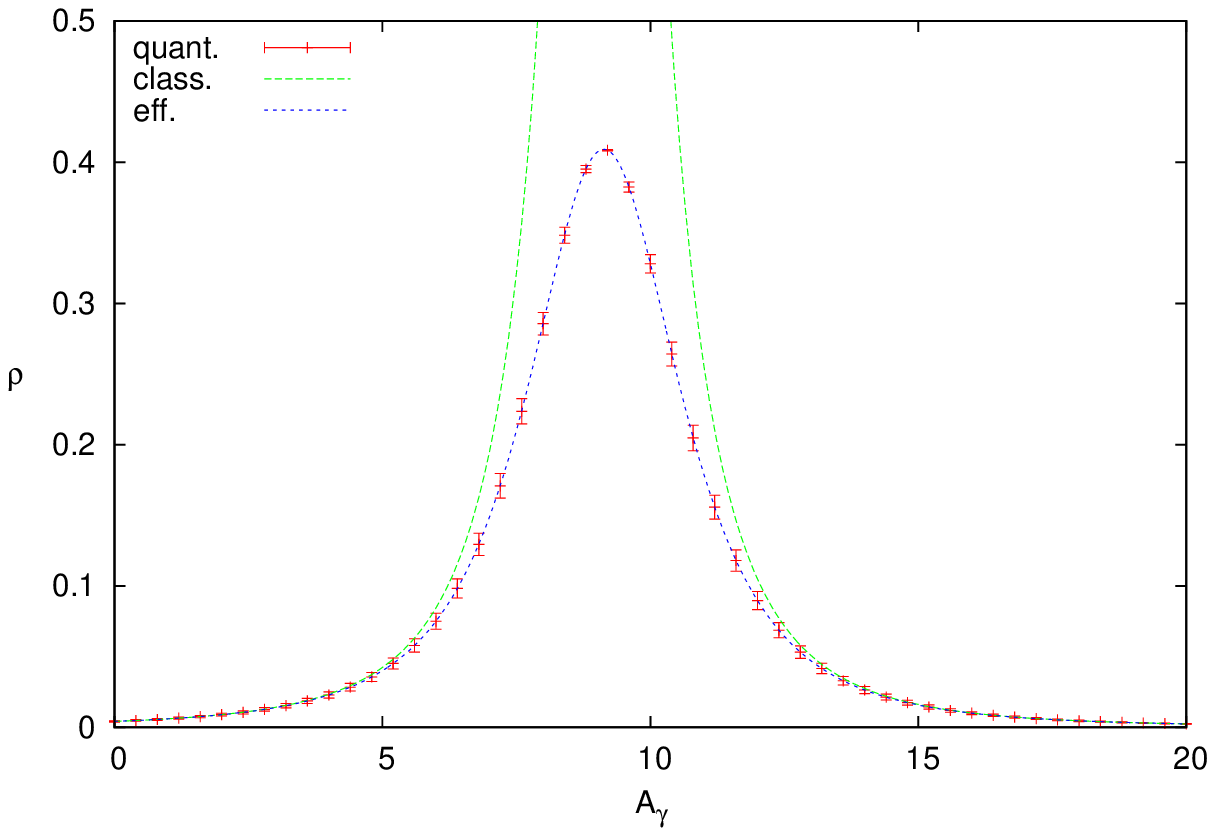}}
  \caption{The expectation values of the observables
		$\wh{x}|_{\Ag} = \wh{|\nu|^{1/3}_{\Ag}}$ and $\wh{\rho}|_{\Ag}$ evaluated
		for the	wave packet presented in Fig.~\ref{fig:dynamics-3d}	are compared
		with the classical trajectories and the evolution 
		predicted by the effective dynamics discussed in Sec.~\ref{s.eff}.}
	\label{fig:dynamics}
\end{figure*}

The results of the analysis (see Figs.~\ref{fig:dynamics-3d} and \ref{fig:dynamics}) 
are fully analogous to the results of studies of the systems with a massless
scalar field \cite{aps-imp} and pressureless dust \cite{hp-dust-LQC}:
\begin{enumerate}
  \item For as long as the energy density [evaluated as the expectation value of 
		$\tilde{\rho}$ in \eqref{eq:rho-g}] is small in comparison to
		$\rho_{\Pl}$, the quantum trajectory follows the classical one.
  \item When the energy density reaches Planck scales the quantum gravity effects 
    	modify the trajectory and lead to a bounce at the critical density $\rho=\rho_c$. 
    	The modification can be heuristically understood as the result of a repulsive
    	gravitational force originating from the underlying fundamental discreteness
		of space-time.
  \item The bounce is a transition epoch deterministically connecting two classical
		epochs of the universe's evolution, when the universe is contracting and expanding
    	respectively.
\end{enumerate}

In the case where the massless scalar field was the matter field,
one of the nice features of the system was an asymptotic preservation of the
semiclassicality. There, the spread of the state in the distant future is strongly 
bounded by its spread in the distant past (and vice versa) through precise triangle
inequalities.  It is therefore natural to test whether an analogous result can also
be obtained in the model studied here. We address this issue in the next section.

\section{Asymptotics of the Dynamics}
\label{s.asym}

In studies of isotropic FLRW cosmologies with a massless scalar field (which
can be used as a clock, just as $\Ag$) it has been shown that in the large $\nu$ limit
all of the eigenfunctions of the LQC evolution operator ---an analog of $\Theta$
\eqref{q-cg-action}--- converge to specific linear combinations of the
eigenfunctions of the evolution operator arising in the Wheeler-DeWitt
(WDW) quantization of the same system \cite{aps-imp}.
This property of the above-mentioned operator permits the modeling of
the global LQC evolution as a certain form of ``scattering'' of WDW universes
(wave packets) by polymeric quantum
geometry effects \cite{kp-scatter}. In a very precise sense, the wave packets 
representing the LQC universe converged in the distant past and future to
``incoming'' (contracting) and ``outgoing'' (expanding) WDW 
wave packets respectively. A detailed study of this picture shows that there
exist rigid relations (in the form of triangle inequalities) between 
the spreads of the LQC wave packet in its distant past and future \cite{kp-scatter}.

This section is dedicated to developing and exploring this same scattering
picture, but in the setting studied in this paper, namely a radiation-dominated
universe. To achieve this, we first construct and study the WDW quantization of 
the radiation-dominated FRLW universe in Sec.~\ref{ss.wdw} and its dynamics, focusing 
in particular on the issue of the uniqueness of the evolution and its relation 
with singularity resolution. Next, in Sec.~\ref{ss.recall} we employ the existence 
of a WDW limit of the LQC evolution operator to construct a scattering picture
and then derive triangle inequalities relevant to the question of 
cosmic recall, analogous to those found in \cite{kp-scatter}. In order to 
provide a stream-lined presentation of this analysis ---including the
improvements necessary to handle Maxwell fields---
Secs.~\ref{ss.wdw} and~\ref{ss.recall} contain only a brief description of
the calculation and results, while the details of the analysis can be found
in Appendices~\ref{app:wdw} and~\ref{app:scatter}.

The triangle inequalities will allow us to provide a rigid bound on how much
the spread of a wave  function can grow from one side of the bounce to the other. 
Thus, this gives an answer to the question of cosmic recall in the following context: 
if the state is semi-classical on one side of the bounce, will it remain so on the 
other side?  In the previous section, numerical studies showed that Gaussian states 
that are initially semi-classical remain sharply peaked throughout their entire 
evolution.  However, determining how the spread of a generic state evolves requires 
stronger and more general methods, of which the scattering picture is a good example.

\subsection{The Wheeler-DeWitt Analog}
	\label{ss.wdw}

The classical cosmological model given in Sec.~\ref{s.class} can be also 
quite easily quantized following the geometrodynamical methods of WDW quantum cosmology, 
by using the standard Schr\"odinger representation rather than the polymer one.
The details of the WDW quantization procedure are presented in Appendix~\ref{app:wdw}. 
Here we briefly summarize the initial assumptions of the procedure and present the 
final result.

Since the WDW analog is to be ultimately used as an approximation to the asymptotics 
of LQC through a scattering picture, it is necessary to ensure that it is as close of an 
analog to our LQC model as is possible.  Therefore, we directly repeat the procedure 
used in Sec.~\ref{s.quant}, in particular by following the Dirac program.  We further extend 
the configuration space to negative $\nu$ to ensure compatibility with LQC, and also choose
same factor-ordering and symmetric superselection sector.

The first step of the Dirac program leads to the kinematical Hilbert space
\begin{equation}
	\ul{\Hil}_{\rm kin} = \ul{\Hil}_{\rm gr} \otimes \Hil_A
	= L^2(\re^+,\rd\nu) \otimes L^2(\re,\rd\Ag) . 
\end{equation}
The basic operators ---quantum counterparts of variables $\nu, b$--- are now multiplication 
and differential operators respectively, and the quantum Hamiltonian constraint takes 
the form
\begin{equation}
	\ul{\Theta}\otimes\Id + \Id\otimes\frac{\partial^2}{\partial\Ag^2} ,
\end{equation}
where the evolution operator $\ul{\Theta}$ is a second-order differential operator
defined on the Schwartz space. The notation is chosen so that all objects in 
WDW theory are represented by same symbols as in LQC, although to differentiate 
them the WDW symbols are underlined.

Unfortunately, $\ul{\Theta}$ is not essentially self-adjoint. Rather, it admits a 
$U(1)$ family of self-adjoint extensions labeled by the parameter $\beta\in [0,\pi)$.
The essential part of the spectrum of each extended operator $\ul{\Theta}_{\beta}$ is 
$\re^+\cup \{0\}$ and is absolutely continuous. 
The physical sector of the theory (identified by group averaging) consists of 
states described by the wave functions
\begin{equation}
	\ul{\Psi}(\nu,\Ag) = \int_{\re^+} \rd k \tilde{\Psi}(k) e_{\beta,k}(\nu) e^{i\omega(k)\Ag} ,
\end{equation}
where the spectral profiles $\tilde{\Psi}\in L^2(\re^+,\rd k)$, the frequency 
$\omega(k)\propto k$ is defined in \eqref{wdw-eigvalue} and the normalized generalized 
basis functions $e_{\beta,k}$ have the form of the standing waves
\begin{equation}\label{eq:wdw-basis}
	e_{\beta,k} (\nu) = \frac{|\nu|^{-1/3}}{\sqrt{6\pi}}\cos[k|\nu|^{1/3}
		+ \ul\varphi(\beta,k)]
\end{equation}
and the extension-dependent phase shift is [see \eqref{eq:wdw-labels}]
\be 
\ul\varphi(\beta,k):=\arctan [\tan(\beta)/k].
\ee

To study the dynamics of this WDW quantum theory, we can use either $(a)$ 
the deparametrization procedure where the evolution is provided by the map
$\re\ni \Ag \mapsto \Psi(\cdot,\Ag) \in \Hil_{\gr}$ [where the basis for each
self-adjoint extension is given in \eqref{eq:wdw-basis} for $\Ag=0$] and
the self-adjoint operators $\wh{\ul{O}}$ on $\Hil_{\gr}$ are the physical
observables, or $(b)$ the relational observables picture, where the
Dirac observables (parametrized by $T$) are the families of operators acting
on $\Hil_{\beta}$ ---where $\beta \in [0, \pi)$ denotes
the one-parameter family of self-adjoint extensions---
as
\begin{equation}
	\wh{\ul{O}}_{T} : \Psi(\nu,\Ag) \mapsto 
	e^{i\sqrt{\ul\Theta_{\beta}}(\Ag-T)} \wh{\ul{O}} \psi(\nu,\Ag=T) .
\end{equation}
In this case, for $\ul\Theta_\beta$, the two approaches are equivalent.

Here, for concreteness, we shall use the deparametrization procedure.
As a convenient set of observables $\wh{\ul{O}}$ we choose
\begin{equation} 
  \uwPg = (-i\hbar/3)\partial_{\Ag}, \quad 
  \wh{\ul{x}} = \wh{|\nu|^{1/3}}.
\end{equation}
Note here that $x$ is positive-definite.
While $\uwPg$ is a constant of motion, the evolution of $\wh{\ul{x}}$
is non-trivial. It is easy to see that the evolution of the quantum universe
is given by an incoming Klein-Gordon wave packet corresponding to the
``pre-singularity'' contracting universe, which upon approaching $x=0$ is
reflected (with a phase shift that depends on the self-adjoint extension and also
$k$) back into an outgoing wave packet corresponding to the ``post-singularity''
expanding universe.  Up to the dispersion of the wave packet, the quantum evolution
follows the (extended) classical trajectory \eqref{c-sol-grav}.

The exact form of the reflected wave packet depends on the chosen self-adjoint
extension, and therefore knowledge of the extension label $\beta$ (which is
equivalent to specifying boundary conditions at $x=0$) is necessary ---in
addition to knowing the initial state--- to uniquely determine the evolution.
In particular, $\beta=0$ corresponds to a simple reflection (as would occur off
an infinite potential barrier) and $\beta=\pi/2$ corresponds to a reflection with 
phase rotation $\pi$ (i.e., sign change). However, all the values $\beta\in[0,\pi)$ 
are allowed and they correspond to the reflective conditions with a particular phase
shift%
\footnote{Note that the choice of $\beta$, 
which determines the reflection conditions of the 
eigenstates from the incoming to the outgoing mode, 
is completely unrelated to the choice of the symmetric 
superselection sector $\ul\psi(\nu) = \ul\psi(-\nu)$.}.

The fact that the choice of the self-adjoint extension $\beta$ affects the
quantum dynamics shows that the singularity is not resolved in the WDW theory.
This is because the choice of $\beta$ is equivalent to setting boundary conditions
at the singularity in order to evolve through the singularity ``by hand''; for
further discussion on this point, see \cite{hp-dust-LQC}.

\subsection{Cosmic Recall}
\label{ss.recall}

The dynamics of this WDW model can be used to accurately describe the asymptotic
dynamics in the distant future and past of the LQC model studied here, as well as provide 
the relation between these epochs via the application of the scattering picture originally 
introduced in \cite{kp-scatter}. The construction of the scattering picture for the
model considered here, including the necessary extensions in order to handle a Maxwell
field, is given in detail in Appendix.~\ref{app:scatter}.  The key points are the following:
\begin{enumerate}
	\item For any localized%
		\footnote{Here by localized we mean a state for which the
			uncertainties of both the observables $\wPg$ and 
			$\wh x_{\Ag}$ are finite in either the distant past or future.}
		LQC physical state $|\Psi\rangle$ there exist two unique WDW physical states 
		$|\Psi_{\rm in}\rangle$ and $|\Psi_{\rm out}\rangle$ such that their dynamics 
		in the distant past and future respectively converge (in the sense of expectation 
		values and dispersions of $\wh x_{\Ag}$) with that of $|\Psi\rangle$, that is
		\begin{subequations}\label{eq:lqc-wdw-obs-main}\begin{align}
			\lim_{\Ag\to\pm\infty} \left[ 
				\langle {\Psi} | \wh{x}_{\Ag} | {\Psi} \rangle 
				- \langle \ul\Psi_{{\rm in}/{\rm out}} | \wh{x}_{\Ag} | 
				\ul\Psi_{{\rm in}/{\rm out}} \rangle_{\rm WDW} \right]
			&=0 , 
			\\
%
			\lim_{\Ag\to\pm\infty} \left[ 
			\langle {\Psi} | \Delta\wh{x}_{\Ag} | {\Psi} \rangle 
			 - \langle \ul\Psi_{{\rm in}/{\rm out}} | \Delta\wh{x}_{\Ag} | 
				\ul\Psi_{{\rm in}/{\rm out}} \rangle_{\rm WDW} \right]
			&=0 ,
		\end{align}\end{subequations}
		and the expectation values and dispersions of $\wPg$ of all three states agree.
	\item The distant past and distant future dispersions of the WDW states are related via 
		the triangle inequality
		\begin{equation}\label{eq:LQC-triangle-main}
			\lim_{\Ag\to\infty} \langle {\Psi} | \Delta\wh{x}_{\Ag} | {\Psi} \rangle  
			\leq 
			\lim_{\Ag\to-\infty} \langle {\Psi} | \Delta\wh{x}_{\Ag} | {\Psi} \rangle 
			+ 2 \langle \Psi | \Delta(\partial_k \vp(k)) | \Psi \rangle .
		\end{equation}
		where the function $\vp(k)$ is the phase shift of the leading-order large $\nu$
		limit of the basis functions $e_{\beta,k}$
		\begin{equation}\label{eq:conv-1}
			e_{\beta,k}(\nu) 
			= \frac{\sqrt{2}}{\sqrt{3\pi}|\nu|^{1/3}} \cos[ k|\nu|^{1/3} + \vp(k) ] 
				+ O(\nu^{-4/3}) .
		\end{equation} 
\end{enumerate}
In order to turn \eqref{eq:LQC-triangle-main} into a useful relation 
(analogous to the one found in \cite{kp-scatter}) we must express 
$\partial_k \vp(k)$ in terms of physically meaningful observables, or 
at least derive an upper bound.
In order to derive an upper bound, it is necessary to resort to numerical analysis, as
detailed in Appendix~\ref{app:triangle}.

The resulting bound ---valid for all superselection sector labels $\epsilon$
and for states supported outside of the interval $[0,k_\star]$ with $k_\star 
\approx 0.15$--- is
\begin{equation}\label{eq:d2scatter-main}
  |\sqrt{k}\partial_k^2 \vp(k)| \leq A/2 , 
\end{equation}
where $A= -0.789\pm 0.005$. Note that the domain in $k$ where the bound holds
is stronger for certain values of $\epsilon$ (see again Appendix~\ref{app:triangle}
for details), and in particular for $\epsilon=0$ the bound is valid for states supported
on $k \in \re^+$.

As a direct consequence, within the domain of validity given above, it is 
possible to rewrite \eqref{eq:LQC-triangle-main} as
\begin{equation}\label{eq:LQC-triangle-final-main}
  \lim_{\Ag\to\infty} \langle {\Psi} | \Delta\wh{x}_{\Ag} | {\Psi} \rangle  
  \leq 
  \lim_{\Ag\to-\infty} \langle {\Psi} | \Delta\wh{x}_{\Ag} | {\Psi} \rangle 
  + A \langle \Psi|\Delta \sqrt{k}|\Psi\rangle ,
\end{equation}
where the only observable besides $\wh x_{\Ag}$ is $\sqrt{k}$, which is proportional to
$\wPg{}^{1/2}$.

This triangle inequality gives a bound on how much the spread in $\wh x_{\Ag}$
can grow from the pre-bounce branch to the post-bounce branch.  This shows the presence
of cosmic recall: a moderately sharply-peaked state in the contracting branch cannot
become wildly quantum in the expanding branch, and vice versa.

\section{Effective Theory}
\label{s.eff}

An interesting and very useful result of LQC is that a classical theory,
obtained by implementing the regularization of the Hamiltonian constraint
but without quantizing it, describes the quantum dynamics of semi-classical states
to a very good degree of precision (with an error well below the
dispersion of the state) in many scenarios
\cite{aps-imp,apsv-spher,bp-negL,*pa-posL,t-eff,*sv-eff,rwe-eff}%
\footnote{Although it is easy to construct the effective dynamics heuristically
	in many contexts of LQC ---including the LQC of isotropic space-times
	with a Schr\"odinger quantization of the matter sector, as is being studied
	here--- one cannot take this heuristic construction for granted in
	more complicated cosmologies.  The details of the formulation of the
	quantum theory, neglected at the effective level, \emph{do significantly
	affect} the genuine quantum dynamics and the existence of semi-classical
	sectors of the theory. For more details, see for example the discussion
	in \cite{kp-polysc}.
	Whether the domain of applicability of the effective dynamics in LQC
	includes more generic cosmologies than isotropic space-times with a Schr\"odinger
	quantization of the matter sector has not yet been determined.
}.
A systematic way to obtain this classical Hamiltonian 
constraint ---called the \emph{effective Hamiltonian constraint}---
is by replacing the shift operators and powers of volume by 
their respective expectation values. The dynamics the effective
Hamiltonian constraint generates are known as the \emph{effective
dynamics of LQC} and have been used extensively.

The effective equations are expected to provide an excellent approximation
to the full quantum dynamics for those states that are sharply peaked.  It has been
shown that, for non-compact space-times (and also compact space-times whose spatial
volume remains much larger than $\lp^3$ at all times), a state which is initially
sufficiently sharply peaked in the semi-classical limit will remain sharply peaked
throughout its entire evolution, including at and around the bounce point where quantum
gravity effects are strongest \cite{rwe-eff}.  As the space-time of interest in
this case is non-compact, the effective equations will indeed provide an excellent
approximation to the full LQC dynamics of sharply peaked states throughout their
entire evolution.  This has been verified numerically for the radiation-dominated
cosmology studied here, and the strong agreement between the effective equations and
the full LQC dynamics can be seen in Fig.~\ref{fig:dynamics}.

Following the procedure described above, the effective Hamiltonian constraint 
(in conformal time $\eta$) takes the form
\be
 \mC_H^{\rm eff}(\eta)  = -\frac{3\pi G\hbar^2}{2\alpha^{2/3}}|\nu|^{4/3}\sin^2 b
 + \f{3}{2} \Pg^2 = 0 ,
\ee
where it is possible to restrict our attention to the sector $\nu > 0$ without any
loss of generality. One of the immediate consequences following from this
equation is that the (classical) matter energy density, originally given 
by Eq.~\eqref{rho} takes the form
\be \label{eff-bound}
\rho = \frac{3}{8 \pi G \ga^2 \Delta} \sin^2 b
\leq \f{3}{8 \pi G \ga^2 \Delta} = \rho_c,
\ee
and thus $\rho$ is bounded above by the critical energy density, just
as in the quantum theory.

The equations of motion generated by $\mC_H^{\rm eff}$ are (in the
sector $\nu > 0$)
\begin{subequations}\begin{align}
	\f{\dd \nu}{\dd \eta} &= \f{6 \pi G \hbar}{\al^{2/3}} \nu^{4/3}  \sin b \, \cos b, &
	\f{\dd b}{\dd \eta} &= -\f{4 \pi G \hbar}{\al^{2/3}} \nu^{1/3} \sin^2 b, \\
	\f{\dd \Ag}{\dd \eta} &= \, \Pg, &
	\f{\dd \Pg}{\dd \eta} &= 0.
\end{align}\end{subequations}
These equations are clearly equivalent to the classical ones in Eq.\ \eqref{eq:eom}
in the classical limit of $b$ being small.

As in the classical theory, $\Pg$ is a constant of the motion and $\Ag$
increases linearly with respect to the conformal time $\eta$.  However,
the solution for $\nu(\eta)$ is now a hypergeometric function and therefore
it is a little harder to see how the quantum corrections modify
the classical dynamics.

In order to clarify this point, it is possible to work in proper time $t$ ($N = 1$)
---rather than the conformal time used above--- as in this case the solution for $\nu(t)$
is much simpler.  For $N = 1$, the effective Hamiltonian constraint (already studied in
\cite{awe-entr}) becomes
\be
\mC_H^{\rm eff}(t) = -\f{3 \pi G \hbar^2}{2 \al} \, \nu \sin^2 b
+ \f{3 \, \Pg^2}{2 (\al \nu)^{1/3}} \approx 0,
\ee
and the equations of motion in proper time are
\begin{subequations}\label{eff-pdot}\begin{align} 
	\f{\dd \nu}{\dd t} &= \f{6 \pi G \hbar}{\al}  \, \nu \sin b \, \cos b, &
	\f{\dd b}{\dd t} &= -\f{3 \pi G \hbar}{\al} \sin^2 b - \f{\Pg^2}{\hbar \al^{1/3} \nu^{4/3}}, \\
	\f{\dd \Ag}{\dd t} &= \, \f{\Pg}{(\al \nu)^{1/3}}, &
	\f{\dd \Pg}{\dd t} &= 0.
\end{align}\end{subequations}
Once again, $\Pg$ is a constant of the motion, but now $\Ag$ is more difficult
to solve for.  However, this choice makes it much easier to solve for $\nu$.
By solving for $\sin^2 b$ via the Hamiltonian constraint, the $\sin b \, \cos b$
term can be replaced by a function of $\nu$ and $\Pg$ which is easy to integrate
as $\Pg$ is constant.  The result is
\be \label{eff-p(t)}
\nu(t) = \f{1}{\al} \left( 16 \pi G \, \Pg^2 \, (t - t_o)^2
+ \f{3 \Pg^2}{2 \rho_c} \right)^{3/4},
\ee
where $t_o$ is a constant of integration. Recalling that the scale factor is
given by $a = (\al \nu)^{1/3}$, this solution shows that for $t$ far away
from $t_o$, the solution approaches the classical trajectory
$a(t) \sim \sqrt{|t - t_o|}$, either for a contracting universe ($t < t_o$)
or an expanding universe ($t > t_o$).  However, there is a major deviation from the
classical solution near $t = t_o$ where there is a bounce in the effective
solution, providing a bridge between the contracting and expanding classical
solutions.  This bounce is what allows the effective solution to avoid the
classical singularity.

As mentioned above, when using proper time $t$, it is much harder to solve for
$\Ag(t)$ (which is a hypergeometric function now), but nonetheless it is possible
examine the dynamics of the matter field by looking at the dynamics of the
energy density and the pressure, which are still determined by the relations
Eqs.\ \eqref{rho} and \eqref{pressure}, where $\nu(t)$ is now given by
Eq.\ \eqref{eff-p(t)}.

Note that the bound on the energy density obtained in Eq.\ \eqref{eff-bound} holds
no matter the choice of the time variable.  In the case of proper time, it is easy
to check that this bound is saturated at $t = t_o$ and that at times far away from
$t_o$, $\rho \ll \rho_c$.

\section{Discussion}
\label{s.dis}

In this paper, we tackled the problem of providing a consistent quantization in the
LQC framework of a flat FLRW universe filled with radiation. This was achieved by 
choosing the matter content to be three copies of Maxwell fields as a toy model
for a photon gas.  For the sake of simplicity, we took the three vector potentials
to be homogeneous, orthogonal and to have equal amplitudes ---the most straightforward
such system that can be coupled to an isotropic geometry.

This particular choice of matter is especially interesting for quantum cosmology
as high temperatures are expected in the very early universe, and the basic
thermodynamic properties at high temperatures relevant for the cosmology of
thermalized matter fields can be modelled by a radiation
gas such as the one studied in this paper. Thus, a radiation-dominated FLRW
loop quantum cosmology is of great relevance for the study of the very early
universe.

An important difference of these studies with respect to previous works in LQC
is that so far the matter content considered in LQC in a genuinely quantum
treatment has never actually been a microscopic field observed in nature: the massless 
scalar field studied in \cite{aps-imp} is neither predicted by particle theories nor 
observed whereas the pressureless dust of \cite{hp-dust-LQC} is observed only as 
a large-scale phenomenon, due to the averaged behaviour of matter fields over
super-galactic scales in cosmology and at micrometer scales in astrophysics.
Here by composing the matter content out of Maxwell fields, we have
incorporated for the first time in LQC a fundamental matter field that
is known to exist.

That being said, it is important to keep in mind that the matter field
considered here is not a rigorous simulation of thermal radiation with a large 
population of fields whose energies follow the Bose-Einstein distribution.
Instead, it is a toy model consisting of three orthogonal homogeneous Maxwell vector
potentials with equal amplitudes. While this is a particularly nice model due 
to its simplicity, it is essential to keep in mind the limitations of such a na\"ive
framework.

This system was quantized within the LQC framework following the standard hybrid
approach, namely a polymer quantization for geometrical degrees of freedom and the
standard Schr\"odinger quantization for the matter sector.
The physical evolution of the quantum system is defined through the set of partial 
observables parametrized by the amplitude of the vector potential, which plays 
the role of an internal clock. The resulting dynamics were studied numerically,
showing that the quantum dynamics are qualitatively similar to the dynamics for
a matter content given by a massless scalar \cite{aps-imp} or a pressureless dust
field \cite{hp-dust-LQC}.  In addition, sharply peaked semi-classical states remain
sharply peaked throughout their evolution, and the global evolution picture features
two classical epochs, one each of contraction and expansion ---where the dynamics follow 
to great precision the predictions of general relativity--- connected deterministically 
by the quantum bounce. The matter energy density remains bounded above with the same
upper bound found in other contexts, $\rho_c = 3/ (8 \pi \ga^2 \Delta G)$.
Furthermore, again as has been done for other matter fields in LQC,
it was possible to define an effective classical description of the system, which
accurately mimicks the genuine quantum evolution for sharply peaked states. 
Finally, the scattering picture of the global evolution can be used in order to
derive strong triangle inequalities between the dispersions of the wave packet 
in the distant past and future, which demonstrates the preservation of the
semi-classicality of the state across the bounce. 

The analytical and numerical findings presented here, together with those of
previous works on isotropic LQC, provide a strong indication that the global evolution
picture and the preservation of semi-classicality across the bounce are generic
features of FLRW universe in LQC: they are independent of the matter content.
Thus, the current results are an additional confirmation of the robustness of
the main results of LQC reported in the literature.

The model studied here also has a particularly interesting property: the big-bang
(or big-crunch) singularity is reached within finite emergent time (this is also the
case for a dust-dominated Friedmann cosmology, but not for a massless
scalar field). This fact makes it possible to directly address and compare the
issue of singularity resolution in LQC and in geometrodynamics for this cosmology.
Indeed, for the Wheeler-DeWitt quantization of this system: $(i)$ the singularity
is reached within the precision set by the wave packet dispersion, and $(ii)$ the
multitude  of self-adjoint extensions implies that additional boundary data
is needed at the singularity in order to deterministically evolve the wave packet
past the point where zero-volume states are reached. These properties amount to
the conclusion that in the WDW quantization the singularity is not resolved.
This is in stark contrast to the results of the loop quantization, where the
unitary evolution is unique and there exists a dynamical minimal volume (proportional 
to $\langle\Pg\rangle^{3/2} \, G^{3} \hbar^{9/4}$, which for semi-classical states is
much greater than the dispersion of the wave packet).  These two results show that
the classical singularity is dynamically resolved in LQC.

On the other hand, note that if a positive cosmological constant is added, the
low curvature dynamics of the cosmology will be similar to that of the FLRW space-time
with a massless scalar field \cite{pa-posL}: the vector potential used here as
a relational clock will become frozen when the cosmological constant dominates the
dynamics, and the infinite volume of the universe will be reached within a finite time.
At the quantum level this implies the non-uniqueness of the unitary evolution
(i.e., there exist many self-adjoint extensions of the evolution generator),
and therefore it is necessary to provide additional data at the boundary $\nu=\infty$ to
evolve the wave function beyond this point.

A last important point is that the LQC FLRW cosmology with Maxwell fields is
significantly more difficult from a technical standpoint than the case of
a massless scalar field. Most notably, the slower rate of convergence of the LQC
evolution operator basis elements to their WDW analogs requires an improvement in
the numerical techniques involved in the analysis, in particular incorporating
higher order LQC corrections to the WDW basis elements.  Since this slower
convergence is expected to be a feature of generic matter content, the improvement
of the numerical treatment that has been presented here is an important further
development of LQC.

\acknowledgments

We would like to thank Jorge Pullin for helpful discussions as well as Brajesh
Gupt and Miguel Megevand for the use of their \texttt{gnuplot} script to generate
Fig.~\ref{fig:dynamics-3d}.  RP would like to thank Carlo Rovelli for his kind
hospitality at the Centre de Physique Th\'eorique (CPT).
This work was supported in part by Le Fonds qu\'eb\'ecois de la recherche sur la nature
et les technologies, the Spanish MICINN grant no FIS2011- 30145-C03-02, the National
Center for Science (NCN) of Poland grants 2012/05/E/ST2/03308 and 2011/02/A/ST2/00300,
the Chilean FONDECYT regular grant no 1140335, as well as a grant from the John Templeton
Foundation. The opinions expressed in this publication are those of the authors and do
not necessarily reflect the views of the John Templeton Foundation. TP further acknowledges
the financial support of UNAB via internal project DI-562-14/R.
The numerical simulations have been performed with the use of the \emph{Numerical LQC} 
library currently developed by T.~Paw{\l}owski, D.~Mart\'in-de Blas and J.~Olmedo. 
The developers thank the Department of Mathematics and Statistics of the Univerity of
New Brunswick (Canada) for hosting the repository of this library in the period 2011-2013. 
The bulk of the computations have been performed on the computational cluster
``Kruk'' at the Institute of Theoretical Physics of Warsaw University (Poland).
The figures have been prepared with the use of the \texttt{gnuplot} software.
The asymptotic expansion of the matrix $M_k$ defined in \eqref{eq:iter-dec} was
determined with the use of the \texttt{Mathematica} symbolic math software.

\appendix

\section{The Wheeler-DeWitt Analog}
	\label{app:wdw}

In this Appendix we discuss in detail the construction and properties of the 
WDW quantization of the radiation-dominated FRLW universe --- the geometrodynamical 
analog of the model studied in the main body of the paper. The procedure is as follows: 
First, in App.~\ref{app.wdw} we repeat the first two steps of the Dirac program in 
the context of the WDW quantization, arriving to the dynamics picture, where the 
evolution is generated by an evolution operator analogous to \eqref{q-cg-action}.
The properties of this operator are next analyzed in App.~\ref{app.sadj}, where all 
of its self-adjoint extensions are identified. This material provides tools for the 
analysis of the dynamics of this WDW model presented in Sec.~\ref{ss.wdw}.

\subsection{The Wheeler-DeWitt Equation}
\label{app.wdw}

In a Wheeler-DeWitt quantization, following from geometrodynamics, one quantizes 
the geometry phase variables using the standard Schr\"odinger representation
rather than the polymer one used in LQC and LQG. The treatment is thus very similar
to the standard textbook procedure, although there are some differences due to the 
cosmological nature of the considered system.

In order to be able to compare the results of this quantization with LQC one 
should proceed in a way as similar as possible to the latter, in particular 
by choosing the same variables. However, in order to demonstrate the qualitative 
properties of the quantum system it is better to start with the original variables
$(a,\pi_{(a)})$ used in \eqref{eq:geom-ham}.

Before proceeding, we have to note that in standard cosmology $a$ being a scale 
factor is positive definite. Since geometrodynamics does not involve triads there 
is no natural reason for equipping it with an orientation. As a consequence the 
gravitational part of the classical phase space is $\re^+\times\re$.

This fact has a critical consequence for the quantization. By choosing the lapse 
$N = a$ (as in LQC) we arrive to the constraint
\begin{equation}\label{eq:constr-ap}
	N \mC_H = - \frac{\pi G}{3} \pi_{(a)}^2 + \frac{3}{2}\Pg^2 , 
\end{equation}
which upon a Schr\"odigner quantization is equivalent to a \emph{Klein-Gordon 
equation on a half-line}%
\footnote{Since the system is a simplification of general
	relativity that is not well-defined on $a=0$, one cannot
	implement any potential barrier there.},
with $a$ as the dynamical variable and $\Ag$ time.  This system can be described
analogously to the Example~$2$ in Section~X.1 of \cite{fa}.  In this case, the analog
of the evolution operator $\ul\Theta$ in \eqref{eq:KG-LQC} (playing the role of
the Hamiltonian) admits a $1$-parameter family of self-adjoint extensions, each extension
corresponding to different reflective boundary conditions at $a=0$. We expect similar
results when using the variables distinguished by LQC.

Let us now perform the quantization in detail using the same variables and operator 
ordering choices as in the LQC quantization. We start with the classical phase space now 
coordinatized by the variables $(\nu,b,\Ag,\Pg)$ specified in \eqref{eq:poisson-matt} 
and \eqref{eq:var-geom}. The main difference with respect to the treatment above is 
the fact that now (following LQC where triads play a crucial role) we equip the variable
$\nu$ with orientation, thus arriving to the classical phase space consisting of two 
copies of the ``purely geometrodynamical'' phase space connected at the $\nu=0$ surface.

By implementing the Schrodinger quantization we arrive to the kinematical Hilbert
space of the gravitational sector which is the space of square-integrable functions
(the Lebesgue space) with respect to the measure $\dd \nu$. As in the case of LQC
the parity invariance of the theory allows us to choose the superselection sector
of symmetric states $\ul\Psi(\nu)=\ul\Psi(-\nu)$ and work within this sector only. 
The variables $\nu$ and $b$ are promoted to operators with the action
\begin{subequations}\label{eq:wdw-op-geom}\begin{align}
	\wh \nu \, \ul\Psi(\nu) &= v \, \ul\Psi(\nu), &
	\wh b \, \ul\Psi(\nu) &= 2 i \frac{\partial}{\partial \nu} \, \ul\Psi(\nu).
\end{align}\end{subequations}
We recall that the notation is chosen so that all objects in WDW 
theory are represented by same symbols as in LQC, although to differentiate 
them the WDW symbols are underlined.

Replacing the basic variables in the classical Hamiltonian constraint 
\eqref{c-ham} by the operators \eqref{eq:op-matt} and \eqref{eq:wdw-op-geom}, and 
choosing a factor-ordering equivalent to the one used in \eqref{q-cg-action},
gives the Wheeler-DeWitt quantum Hamiltonian constraint,
\be \label{eq:const-wdw}
-\f{\partial^2}{\partial \Ag^2} \ul\Psi(\nu, \Ag)
= \ul\Theta \, \ul\Psi(\nu, \Ag),
\ee
where the WDW evolution operator is
\begin{equation} \label{wdw-op}
	\ul\Theta = 18 \f{(2 \pi \ga \sqrt\Delta \lp^3)^{1/3}} {\ga \lp \sqrt\Delta \hbar}\,
	|\nu|^{1/3} \wh{D}|\nu|^{2/3}\wh{D}|\nu|^{1/3},
\end{equation}
with the operator $\wh{D}$ defined as
\begin{equation}\label{eq:D-def}
  \wh{D} = \frac{i}{2}\left[ \sgn(\nu)\f{\partial}{\partial \nu}
		+ \f{\partial}{\partial \nu}\sgn(\nu) \right] .
\end{equation}
This particular form of $\wh{D}$ is a consequence of implementing the 
factor-ordering choices made for the LQC Hamiltonian constraint in this 
paper (called the sMMO factor-ordering in \cite{mop-presc}). On the open domain 
disjoint from $\nu=0$, $\ul\Theta$ is equivalent to an operator \eqref{wdw-op} with 
$\wh{D}$ replaced with $i\partial_{\nu}$, however due to presence of $\sgn(\nu)$
special care is required at $\nu=0$. In particular one has to restrict the domain 
of $\ul\Theta$ [for which one would usually choose the Schwartz space 
$\mathcal{S}(\re)$] setting
\begin{equation}\label{eq:wdw-dom} 
	\mathcal{D}(\ul\Theta) = \{ \ul\psi \in\mathcal{S}(\re):\ 
	\ul\psi(\nu) = \ul\psi(-\nu) 
	\wedge \ul\psi(\nu=0)=0 \} .
\end{equation}
From the symmetry and differentiability of $\ul\psi$, it follows that
\begin{equation}\label{wdw:dom-cond}
	\partial_{\nu}|\nu|^{1/3}\ul\psi(\nu)|_{\nu=0} = 0 .
\end{equation}
The $\ul\Theta$ operator is symmetric and non-negative definite on this domain.

Its symmetric eigenfunctions 
\be
\ul\Theta \, \ul e_k(\nu) = \ul \omega^2_k \, \ul e_k(\nu)
\ee
correspond to positive real eigenvalues, and are 
Dirac delta normalizable; imposing the normalization to be
$\braket{\ul e_{k}}{\ul e_{k'}} = \delta(k-k')$, the
eigenfunctions are
\be \label{wdw-eigenf}
\ul e_k(\nu) = \f{|\nu|^{-1/3}}{\sqrt{6\pi}} e^{i k |\nu|^{1/3}},
\ee
where the label $k$ (an analog of the wave number) spans the entire 
real line. The relation between the eigenvalues of $\ul\Theta$ and
the $k$ labels (the analog of the dispersion relation) is%
\footnote{There is a freedom in the definition of $k$: we could instead choose
	$\ul e'_k(\nu) = {\nu^{-1/3}} \sqrt{\ell/6\pi} \exp[i \ell k \nu^{1/3}]$,
	in which case $(\ul\omega')^2_k = \ell^2 \ul\omega^2_k$.  The 
	(arbitrary) length scale $\ell$ can clearly be absorbed into the 
	definition of $k$ and this is what is done here.
}
\be \label{wdw-eigvalue}
\ul\omega^2_k = 2 \f{(2 \pi \ga \sqrt\Delta)^{1/3}}
{\ga \sqrt\Delta \hbar} k^2.
\ee

\subsection{Self-Adjoint Extensions}
\label{app.sadj}

It is easy to see by inspection that the Wheeler-DeWitt evolution operator  
\eqref{wdw-op} is symmetric in its domain $\mathcal D(\ul \Theta)\equiv %
\mathcal{S}(\re)$ and that its spectrum is real and non-negative. However in order 
to generate unique unitary evolution, the operator has to satisfy the stronger 
requirement of being essentially self-adjoint. Here we verify this property by 
studying its deficiency spaces.

The existence and uniqueness of self-adjoint extensions
to $\ul\Theta$ is particularly important in the context of singularity 
resolution, as it answers the question whether unitary evolution of the state 
is possible and whether any additional data is needed at the former classical 
singularity to determine the evolution uniquely. Once the self-adjoint
extensions are known, it is possible to study the dynamics and determine
whether the singularity is in fact resolved or not. This is done in
Sec.~\ref{ss.wdw}. 

Our first step in determining the self-adjoint extension to $\ul\Theta$ is the 
identification of the deficiency subspaces denoted by $\mathcal K^{\pm} $
that are the spaces of normalizable solutions to the equation
\be\label{eq:defi-eq}
	\ul{\Theta} \, \ul{e}_{\pm i}=\pm i \,\ul{e}_{\pm i}\,,
\ee
i.e., normalizable eigenfunctions with eigenvalues $\pm i$. The above 
equation is easy to solve analytically. Its normalizable solutions are all 
proportional to the two following normalized functions
\be\label{def-eif}
\ul e_{\pm i}(\nu)=\frac{1}{(18 \om_o^2)^{1/4}}\,\frac{1}{|\nu|^{1/3}}\,
e^{-(1\mp i)|\nu|^{1/3} / \sqrt 2 \om_o} \,,
\ee
where
\be\label{eq:wdw-disp}
\om_o^2 = 2 \f{(2 \pi \ga \sqrt\Delta)^{1/3}}{\ga \sqrt\Delta \hbar}.
\ee
There also exists a second family of formal solutions to \eqref{eq:defi-eq} with 
a growing exponential, but it is not normalizable in the kinematical Hilbert 
space and therefore it does not contribute to the deficiency space. 

As a consequence, each of the deficiency spaces $\mathcal K^{\pm}$ is 
one-dimensional:  $\mathcal K^{\pm}=span\{\ul e_{\pm i}(\nu)\}$.
Thus, according to Theorem X.2 in \cite{fa}, $\ul\Theta$
admits many self-adjoint extensions, 
each corresponding to a unitary map $U^\alpha: \mathcal K_+ \to \mathcal K_-$. 
Since ${\rm dim}(\mathcal{K}_+)={\rm dim}(\mathcal{K}_-)=1$ such maps form a
$1$-dimensional family parametrized by $\alpha\in[0,\pi)$, each element being 
$U^\alpha e_+(\nu) = e^{i\alpha} e_-(\nu)$, exactly as expected from the preliminary 
considerations given in Sec.~\ref{ss.wdw}. Each extended 
domain $\mathcal D_{\alpha}(\ul \Theta)$ then takes the form
\begin{equation}\label{eq:wdw-dom-ext}
		\mathcal D_{\alpha}(\ul\Theta) = \{ \ul\psi_{ext}:\ 
		\ul\psi_{ext}(\nu) = \ul\psi(\nu) 
		+ \lambda \ul{e}_{\alpha}(\nu),\ 
		\ul\psi \in \mathcal D(\ul \Theta), \lambda\in\mathbb{C} \},
\end{equation}
where 
\begin{equation}\label{eq:wdw-ea}
	\ul e_{\alpha}(\nu) := \ul{e}_{+i}(\nu) + e^{i\alpha}\,\ul{e}_{-i}(\nu) 
	= \frac{2}{(18 \om_o^2)^{1/4}}\,\frac{1}{|\nu|^{1/3}}\,
		e^{-|\nu|^{1/3} / \sqrt 2 \om_o}
		\cos\left(\frac{1}{\sqrt 2 \om_o} |\nu|^{1/3}-\f{\alpha}{2}\right)\,
		e^{i\f{\alpha}{2}} \,.
\end{equation}
Instead of identifying explicit boundary conditions at $\nu=0$ associated with each 
extension, in this case it is easier to find the extended bases through the orthogonality
requirement. First we note that \eqref{eq:wdw-dom-ext} and \eqref{eq:wdw-ea} imply 
that the basis elements of any self-adjoint extension have to contain balanced
``incoming'' and ``outgoing'' components, that is
\begin{equation}\label{eq:wdw-labels}
	\ul{e}_{\alpha,k} = \frac{1}{\sqrt{3}\pi|\nu|^{1/3}}\cos \big[k|\nu|^{1/3} 
		+ \ul\vp(\alpha,k) \big] .
\end{equation}
The requirement of orthogonality within each extension basis (labelled by $\alpha$) 
implies the selection condition
\begin{equation}\label{eq:wdw-ext-phase}
  \tan[\ul\vp(\alpha,k)] = \frac{\tan[\beta(\alpha)]}{k} ,
\end{equation}
where $\beta: [0,\pi) \to [0,\pi)$ is a bijective function of the extension label 
$\alpha$. From now on we will use $\beta$ as the extension label for technical 
convenience.

Each self-adjoint extension $\ul\Theta_\beta$ of $\ul\Theta$ has a
non-degenerate spectrum of which the non-negative part (being the only 
one contributing to the physical sector)%
\footnote{Note that due to the non-negativity of
$-\del^2 / \del \Ag^2$ on $\mathcal{H}_A$ any a priori contribution from the 
negative part of the spectrum of $\ul\Theta_\beta$ would be removed from the 
physical sector in the process of solving \eqref{eq:const-wdw}
(equivalently $N \mC_H = 0$) through group averaging.}
is absolutely continuous
${\rm Sp}(\ul\Theta_\beta) = \re^+ \cup \{0\}$. The spaces 
of physical states (the positive frequency sector) for each extension are 
\begin{equation}\label{eq:wdw-phys-ext}
  \Hil_{\beta}\ni \Psi(\nu,\Ag) 
  = \int_{\re^+} \rd k \, \tilde{\ul\Psi}(k) \, \ul{e}_{\beta,k}(\nu) \,
  e^{i\omega(k)\Ag} ,
\end{equation}
where $\omega(k)$ is given by \eqref{eq:wdw-disp} and the spectral profile
$\tilde{\ul\Psi}$ is normalizable in $L^2(\re^+,\rd k)$.

\section{The Scattering Picture}
	\label{app:scatter}

Having at our disposal the well-defined WDW theory of App.~\ref{app:wdw},
we can relate the asymptotics of the LQC dynamics to the WDW dynamics using the 
scattering picture introduced in \cite{kp-scatter}. However, due to the more complicated 
(as compared to the case of a space-time with a massless scalar field) structure of the 
WDW theory itself, some improvements have to be made to the method used. In particular, 
it is necessary to introduce the scattering picture already at the level of the WDW theory 
itself, using in the process certain auxiliary structures. This is done in App.~\ref{sss.wdw-scat}. 
These structures will be used in App.~\ref{sss.lqc-scatt} to build in turn the scattering 
picture in LQC.  Finally, the proper WDW limit of the LQC dynamics is presented in 
App.~\ref{sss.wdw-lqc} and from this it is possible to derive useful triangle inequalities
between the dispersions of certain physically interesting observables,
as shown in App.~\ref{app:triangle}.

\subsection{Wheeler-DeWitt Scattering}
\label{sss.wdw-scat}

As each basis element of a physical Hilbert space has the form of a reflected
plane wave, it is natural to split it into the incoming and outgoing components,
\begin{equation}\label{eq:wdw-decomp}
	\ul{e}_{\alpha,k}(\nu) = \frac{1}{\sqrt{2}}
	\left( e^{i \ul\vp(\alpha,k)} \ul{e}^+_k + e^{-i \ul\vp(\alpha,k)} \ul{e}^-_k \right),
\end{equation}
where
\begin{equation}\label{eq:aux-basis}
	\ul{e}^{\pm}_k = \frac{1}{\sqrt{6\pi}|\nu|^{1/3}} e^{\pm ik|\nu|^{1/3}} .
\end{equation}
The terms $\ul{e}^{\pm}_k$ can be thought of as the incoming and outgoing 
plane waves in the auxiliary Hilbert space $\ul{\Hil}_{\rm aux}$ constructed by 
$(i)$ restricting the support of the symmetric wave functions on $\ul\Hil_{\rm phy}$
and $(ii)$ extending it to the (now unphysical) domain $\nu< 0$ by taking the 
following extension of $\ul{e}^{\pm}_k$:
\begin{equation}
  \ul{\tilde{e}}^{\pm}_k = \frac{1}{\sqrt{6\pi}|\nu|^{1/3}} e^{\pm ik\,\sgn(\nu)|\nu|^{1/3}} .
\end{equation}
Given that, one can treat the reflection of the WDW wave packet at the singularity as 
a specific example of a scattering, that is the transition
\begin{equation}\label{eq:scatter-wdw-state}
	\int_{\re^+} \rd k \, \tilde{\Psi}_{\rm in} \, \ul{\tilde{e}}^{+}_k(\nu) \, 
	e^{i\omega\Ag} 
	=: \Psi_{\rm in}(\nu,\Ag) 
	\ \mapsto\ \Psi_{\rm out}(\nu,\Ag)
	:= \int_{\re^+} \rd k \, \tilde{\Psi}_{\rm out} \, \ul{\tilde{e}}^{-}_k(\nu) \, 
	e^{i\omega\Ag} .
\end{equation}
The decomposion \eqref{eq:wdw-decomp} implies that 
\begin{equation}\label{eq:wdw-spectr-scatt}
  e^{-i \ul\vp(\alpha,k)} \tilde{\Psi}_{\rm in}(k) 
  = e^{i\ul\vp(\alpha,k)} \tilde{\Psi}_{\rm out}(k) 
  = \tilde{\Psi}(k) ,
\end{equation}
thus the scattering process is described by the density matrix $\wh{\rho}$, where
\begin{equation}
	\rho(k,k') = e^{-2i \ul\vp(\alpha,k)} \delta(k-k') .
\end{equation}

The observables $\uwPg$ and $\wh{\ul{x}}_{\Ag}$
defined on $\ul\Hil_{\rm phy}$ can be transferred in a straightforward way 
to the observables $\wPg$ and $\wh x_{\Ag}$ defined on the auxiliary space 
$\ul\Hil_{\rm aux}$, such that
\be
  \wPg = -\frac{i\hbar}{3}\partial_{\Ag} , \qquad 
  \wh{x}_{\Ag} = \wh{\nu_{\Ag}^{1/3}}.
\ee
Then, for sufficiently localized WDW states, that is states that satisfy
$\langle\Delta\wPg\rangle<\infty$, $\langle\Delta\wh{x}_{\Ag}\rangle<\infty$
(see for example \cite{kp-scatter} for a discussion), we have
\begin{subequations}\label{eq:wdw-aux-obs}\begin{align}
		\lim_{\Ag\to\pm\infty} \left[ 
			\langle {\ul\Psi} | \wh{x}_{\Ag} | {\ul\Psi} \rangle_{\rm WDW} 
			  - \langle \ul\Psi_{{\rm in}/{\rm out}} | \wh{x}_{\Ag} | 
			\ul\Psi_{{\rm in}/{\rm out}} \rangle_{\rm aux} \right]
			&= 0 , 
	\\
		\lim_{\Ag\to\pm\infty} \left[ 
		\langle {\ul\Psi} | \Delta\wh{x}_{\Ag} | {\ul\Psi} \rangle_{\rm WDW} 
		  - \langle \ul\Psi_{{\rm in}/{\rm out}} | \Delta\wh{x}_{\Ag} | 
			\ul\Psi_{{\rm in}/{\rm out}} \rangle_{\rm aux} \right]
		&= 0 ,
\end{align}\end{subequations}
which allow us to find explicit relations between the dispersions of
$\wh{x}_{\Ag}$ in the distant future and past. 
In particular, as the operator $\wh{x}_{\Ag}$ on $\ul\Hil_{\rm aux}$ has the form 
$\wh{x}_{\Ag}=-i\partial_k + i\Ag[\partial_k\omega]\Id$ and the expectation values and 
dispersions of the operator $-i\partial_k$ on $\ul\Psi_{{\rm in}/{\rm out}}$ are
related via
\begin{subequations}\begin{align}
	\langle \ul\Psi_{\rm out}|-i\partial_k|\ul\Psi_{\rm out} \rangle 
  &= \langle \ul\Psi_{\rm in}|-i\partial_k-2[\partial_k \ul\vp]\Id|\ul\Psi_{\rm in} \rangle,
    \\
  \langle \ul\Psi_{\rm out}|\Delta(-i\partial_k)|\ul\Psi_{\rm out} \rangle 
  &= \langle \ul\Psi_{\rm in}|\Delta(-i\partial_k-2[\partial_k \ul\vp]\Id)|\ul\Psi_{\rm in} \rangle,
\end{align}\end{subequations}
we can easily construct (following the derivation in \cite{kp-scatter}) a ``triangle inequality'' 
involving the dispersions
\begin{equation}\label{eq:WDW-triangle}
  \lim_{\Ag\to\infty} \langle {\ul\Psi} | \Delta\wh{x}_{\Ag} | {\ul\Psi} \rangle_{\rm WDW} 
  \leq 
  \lim_{\Ag\to-\infty} \langle {\ul\Psi} | \Delta\wh{x}_{\Ag} | {\ul\Psi} \rangle_{\rm WDW} 
  + 2 \langle \ul\Psi | \Delta(\partial_k \ul\vp) | \ul\Psi \rangle_{\rm WDW}
\end{equation}
where from \eqref{eq:wdw-ext-phase} it follows that for the self-adjoint extension labelled
by $\beta$
\begin{equation}\label{eq:alpha-beta}
  \partial_k \ul\vp = \frac{\tan(\beta)}{k^2+\tan^2(\beta)}\Id . 
\end{equation}
For states sharply peaked in $\Pg$ the dispersion of the operator $\partial_k \ul\vp$ behaves 
approximately like $ \langle \Delta \partial_k \ul\vp \rangle \sim \langle \Pg \rangle^{-3} %
\langle \Delta\Pg \rangle$, thus this inequality shows that there is a strong preservation
of semiclassicality in the process of the transition between the ``incoming'' and ``outgoing''
modes.

\subsection{LQC Scattering}
\label{sss.lqc-scatt}

In the case of flat FRLW cosmologies with a massless scalar field, in the
large $\nu$ limit the eigenfunctions of the LQC evolution operator approach
a particular combination of the eigenfuctions of the WDW evolution operator. This 
permits a description of the global LQC dynamics as the process of a
scattering of WDW wave packets in a way similar to the example described
in Sec.~\ref{sss.wdw-scat}.

For a radiation-dominated FLRW space-time, a precise 
analog of this result may not be possible, as each WDW physical Hilbert space element 
is already a combination of two plane waves. However, the procedure can be
generalized so that, instead of using the WDW basis directly,
one uses the incoming and outgoing components that form a basis on the auxiliary 
space defined in \eqref{eq:aux-basis}.

Therefore, it is necessary to verify the convergence%
\footnote{Here we use the textbook nomenclature where
    $\lim O(f(x))/f(x) <\infty$ and $\lim o(f(x))/f(x) = 0$.}
of the basis elements $e_k$
\be \label{eq:pre-conv}
  e_k(\nu) = f(k) \ul e_k^{+}(\nu) + f(-k) \ul e_{k}^{-}(\nu) 
    + |\ul e_k^{+}(\nu)|o(|\nu|^0) ,
\ee
where $|f(k)| = |f(-k)|$ due to the reality of $e_k$ as solutions to \eqref{eq:eig}.

The expectation that \eqref{eq:pre-conv} holds comes from studying the numerical
evaluation of $e_k$. The convergence to the incoming/outgoing components 
of the WDW wave packets in the distant past/future has also been observed directly at
the level of states. To confirm the validity of \eqref{eq:pre-conv}, we use the
analytic method specified in \cite{kp-pcc}. 
Its core elements are:
\begin{enumerate}
  \item Rewriting the second order iterative relation between consecutive points of the 
    support of $e_k$ in the first order form
    \begin{equation}
      \vec e_k(\nu + 4) = A_k(\nu) \, \vec e_k(\nu),
    \end{equation}
    where
    \begin{subequations}\begin{align}
      \vec e_k(\nu) &= \left( \begin{array}{c}
        e_k(\nu) \\ e_k(\nu - 4)  
			\end{array}\right) \,,  &
      A_k(\nu) &= \left( \begin{array}{cc}
        \f{f_o(\nu) - \omega^2(k)}{f_+(\nu)} & - \f{f_-(\nu)}{f_+(\nu)}  \\
        1 & 0
      \end{array} \right) \,,
    \end{align}\end{subequations}
    with $f_o,f_{\pm}$ specified via \eqref{Theta-op}.
  \item Expressing the values of $e_k$ on each pair of the consecutive points of its support 
    as a linear combination of the WDW components [corresponding to the same $\omega(k)$],
    which in the notation above can be written as
    \begin{subequations}\label{eq:vec-wdw-dec}\begin{align}
      \vec e_k(\nu) &= B_k(\nu) \, \vec \chi_k(\nu), &
			B_k(\nu) &= \left( \begin{array}{cc}
			\ul e^{+}_k(\nu) & \ul e^{-}_{k}(\nu)  \\
			\ul e^{+}_k(\nu - 4) & \ul e^{-}_{k}(\nu - 4)
			\end{array}\right) \,,
    \end{align}\end{subequations}
    where the matrix $B_k(\nu)$ is invertible for sufficiently large $|\nu|$. If the expected 
    convergence \eqref{eq:pre-conv} holds, then the coefficient vector $\chi_k$ has a
		well-defined large $|\nu|$ limit.
  \item Rewriting the eigenvalue problem as a first order iterative equation for the coefficient
    vectors
    \begin{subequations}\label{eq:iter-dec}\begin{align}
       \vec \chi_k &= M_k(\nu) \, \vec \chi_k, &
       M_k(\nu) &:= B_k^{-1}(\nu + 4) \, A(\nu) \, B_k(\nu) .
    \end{align}\end{subequations}
    Then the condition sufficient for the convergence \eqref{eq:pre-conv} is
    \begin{equation}
      M_k(\nu) = \Id + o(\nu^{-1}) ,
    \end{equation}
    and the problem is reduced to probing the asymptotics of $M_k(\nu)$.
\end{enumerate}
Direct inspection shows that 
\begin{equation}
  M_k(\nu) = \Id + O(\nu^{-2}) \ ,
\end{equation} 
which shows that the relation \eqref{eq:pre-conv} indeed holds.
Then, the reality of $e_k$ and the comparison of the normalizations in $\Hil_{\rm phy}$ and 
$\Hil_{\rm aux}$ (see \cite{aps-det,kp-scatter,mop-presc} for details in an analogous 
setting) indicate that \eqref{eq:pre-conv} can be written as%
\footnote{At this point 
  it is not yet obvious if we can relate the normalization of $e_k$ on $\Hil_{\rm phy}$ with 
  the normalization of $e^{i\varphi(k)} \ul{e}^+_k(\nu) + e^{-i\varphi(k)} \ul{e}^-_k(\nu)$
  on $\Hil_{\rm aux}$ that was derived in \cite{kp-scatter} for second order convergence. 
  However, we see from \eqref{eq:lim-imp} that the first subleading term is oscillatory, so
  the decay rate of $\nu^{-1}$ is indeed sufficient for it to not contribute to the
  normalization.
}
\begin{equation}\label{eq:conv-actual}
  e_k(\nu) = e^{i\varphi(k)} \ul{e}^+_k(\nu) + e^{-i\varphi(k)} \ul{e}^-_k(\nu) 
    + |\ul{e}^+_k(\nu)|O(\nu^{-1}) .
\end{equation} 
Note that the convergence is one order weaker than for the case of a massless
scalar field.

While this result is sufficient to construct the scattering picture, for practical 
numerical applications (like evaluating $\lim_{\nu\to\infty} \vec\chi_k(\nu)$, which is
needed for normalization of $e_k$), the convergence is too slow. The rate of
convergence can be improved by replacing the components $\ul{e}^{\pm}_k$ in
\eqref{eq:vec-wdw-dec} with the functions%
\footnote{A systematic procedure to determine this function is
  to calculate the subleading terms order by order by using the constraints that arise from 
  imposing the appropriate level of convergence on the analog $M^{(l)}_k$ of $M_k$ 
  corresponding to given order $l$.
}
\begin{equation}\label{eq:lim-imp}
	\ul e'_k(\nu) = \f{|\nu|^{-1/3}}{\sqrt{6 \pi}} \left[ 1 + \f{k^2}{9 |\nu|^{4/3}}
	+ \f{5 k^4}{162 |\nu|^{8/3}} \right] 
	e^{ i k \left( |\nu|^{1/3}
	- \f{2 k^2}{81 |\nu|} + \f{4}{27 |\nu|^{5/3}} - \f{2 k^4}{945 |\nu|^{7/3}}
	\right)},
\end{equation}
thus constructing the analog $M^{(4)}_k$ of matrix $M_k$ defined in \eqref{eq:iter-dec}.
Direct inspection of the asymptotics of $M^{(4)}_k$ shows that
\begin{equation}
  M^{(4)}_k(\nu) = \Id + O(\nu^{-4}) , 
\end{equation}
which implies that
\begin{equation} \label{eq:conv-imp} 
  e_k(\nu) = e^{i\varphi(k)} \ul{e}'_k(\nu) + e^{-i\varphi(k)} \ul{e}'_{-k}(\nu) 
    + |\ul{e}'_k(\nu)|O(\nu^{-3}) .
\end{equation}

The relation \eqref{eq:conv-actual} allows us to again introduce the scattering picture
as a mapping of the type \eqref{eq:scatter-wdw-state} between the auxiliary states, 
where the spectral profiles of $\ul\Psi_{{\rm in}/{\rm out}}$ are related with 
the LQC spectral profile $\tilde{\Psi}$ \eqref{eq:psi} via
\begin{equation}\label{eq:lqc-spectr-scatt}
  e^{-i\varphi(k)}\tilde{\Psi}_{\rm in}(k) 
  = e^{i\varphi(k)}\tilde{\Psi}_{\rm out}(k) = \tilde{\Psi}(k) ,
\end{equation}  
which gives a scattering matrix of the form
\begin{equation}
  \rho(k,k') = e^{-2i\varphi(k)}\delta(k-k') .
\end{equation}

Due to the oscillatory nature of the first subleading correction in \eqref{eq:conv-actual},
it is possible to relate the distant past/future expectation values
and observables for localized LQC states (defined by the conditions
$\langle\Delta\wPg\rangle<\infty$ and $\langle\Delta\wh{x}_{\Ag}\rangle<\infty$)
with those of the incoming/outgoing auxiliary states
\begin{subequations}\label{eq:lqc-aux-obs}\begin{align}
  \lim_{\Ag\to\pm\infty} \left[ 
    \langle {\Psi} | \wh{x}_{\Ag} | {\Psi} \rangle 
    - \langle \ul\Psi_{{\rm in}/{\rm out}} | \wh{x}_{\Ag} | 
    \ul\Psi_{{\rm in}/{\rm out}} \rangle_{\rm aux} \right]
		&= 0 , 
  \\
  \lim_{\Ag\to\pm\infty} \left[ 
  \langle {\Psi} | \Delta\wh{x}_{\Ag} | {\Psi} \rangle
    - \langle \ul\Psi_{{\rm in}/{\rm out}} | \Delta\wh{x}_{\Ag} | 
    \ul\Psi_{{\rm in}/{\rm out}} \rangle_{\rm aux} \right]
		&= 0 ,
\end{align}\end{subequations}
by adapting the construction given in Appendix~A2 of \cite{kp-scatter}
to this setting.

\subsection{The WDW Limit of LQC Dynamics}
\label{sss.wdw-lqc}

Because the basis $\ul{e}_k$ of the WDW quantum cosmology is slightly more
complicated than in the case of the massless scalar field, it was necessary
to define the scattering of the LQC states by using the auxiliary space and
its basis functions. In consequence, the results given in Sec.~\ref{sss.lqc-scatt}
do not provide a direct relation between the LQC and WDW states.

Fortunately, it is possible
(and easy) to describe the evolution of the WDW state itself as the scattering 
of the auxiliary state. Furthermore, the auxiliary space emerging in the scattering 
picture of LQC state is the same as for WDW.  This allows us to employ the 
auxiliary in/out states as an intermediate providing the relation between the 
LQC and WDW states. Indeed, given an LQC state, the WDW \emph{in} 
(\emph{out}) state is defined by the requirement that the auxiliary \emph{in} (\emph{out}) 
component in the scattering description of that state agrees with the auxiliary 
\emph{in} (\emph{out}) component in the scattering description of the LQC state itself. 

In other words, the relation between the spectral profiles of these states
---given by \eqref{eq:wdw-spectr-scatt} and \eqref{eq:lqc-spectr-scatt}--- takes the
form
\begin{equation}
  e^{i[ \ul\vp(\alpha,k) - \varphi(k)]} \tilde{\ul\Psi}_{\rm in} 
  = e^{-i[\ul\vp(\alpha,k)-\varphi(k)]} \tilde{\ul\Psi}_{\rm out} 
  = \tilde{\Psi}(k) ,
\end{equation}
where $\alpha'(\alpha,k)$ is given by \eqref{eq:wdw-ext-phase} and 
$\tilde{\ul\Psi}_{\rm in}$ and $\tilde{\ul\Psi}_{\rm out}$ are the spectral 
profiles of the WDW in and out states respectively. As a consequence, we can describe 
the global LQC evolution as the \emph{scattering of WDW states}. The scattering 
matrix of this process is given by
\begin{equation}\label{eq:lqc-wdw-rho}
  \rho(k,k') = e^{-2i[\varphi(k)- \ul\varphi(\alpha,k)]} \delta(k-k') .
\end{equation}
It is important to remember that defining this picture requires us to choose one
particular (labeled by $\alpha$) self-adjoint extension of $\ul\Theta$. The scattering 
matrix \eqref{eq:lqc-wdw-rho} depends on this choice.

The relations \eqref{eq:wdw-aux-obs} and \eqref{eq:lqc-aux-obs} allow us to 
provide a relation between the expectation values and dispersions of the $\wh{x}_{\Ag}$ 
operator in the distant future and past,
\begin{subequations}\label{eq:lqc-wdw-obs}\begin{align}
  \lim_{\Ag\to\pm\infty} \left[ 
    \langle {\Psi} | \wh{x}_{\Ag} | {\Psi} \rangle
      - \langle \ul\Psi_{{\rm in}/{\rm out}} | \wh{x}_{\Ag} | 
    \ul\Psi_{{\rm in}/{\rm out}} \rangle_{\rm WDW} \right]
		&= 0 , 
  \\
  \lim_{\Ag\to\pm\infty} \left[ 
  \langle {\Psi} | \Delta\wh{x}_{\Ag} | {\Psi} \rangle
    - \langle \ul\Psi_{{\rm in}/{\rm out}} | \Delta\wh{x}_{\Ag} | 
    \ul\Psi_{{\rm in}/{\rm out}} \rangle_{\rm WDW} \right]
		&= 0 .
\end{align}\end{subequations}
The expectation values and dispersion of the operator $\wPg$ of the in/out WDW 
states are exactly that of the LQC state, since the relations \eqref{eq:wdw-spectr-scatt} 
and \eqref{eq:lqc-spectr-scatt} are only phase rotations.

\subsection{The Triangle Inequality}
	\label{app:triangle}

While in Sec.~\ref{sss.wdw-lqc} we defined a precise description of the global 
evolution of the LQC state as the scattering of certain WDW states, to relate the 
spreads of the LQC state in the distant future and past we will employ the scattering 
picture defined in Sec.~\ref{sss.lqc-scatt} which uses the auxiliary states.
This choice is motivated by the fact that in the auxiliary space the operator 
$\wh{x}_{\Ag}$ has a simple analytical form in the $k$-representation.
Indeed, the kinematical operator (or the physical observable in the deparametrization
picture) is
\begin{equation}
  \wh{x} = -i\partial_k + [\partial_k\omega]\Ag\Id ,
\end{equation}
That, together with \eqref{eq:lqc-aux-obs}, allows us to immediately write down the 
triangle inequality analogous to \eqref{eq:WDW-triangle},
\begin{equation}\label{eq:LQC-triangle}
  \lim_{\Ag\to\infty} \langle {\Psi} | \Delta\wh{x}_{\Ag} | {\Psi} \rangle  
  \leq 
  \lim_{\Ag\to-\infty} \langle {\Psi} | \Delta\wh{x}_{\Ag} | {\Psi} \rangle 
  + 2 \langle \Psi | \Delta[\partial_k \vp(k)] | \Psi \rangle .
\end{equation}

\begin{figure*}[tbh!]
	\subfigure[]{\includegraphics[width=0.49\textwidth]{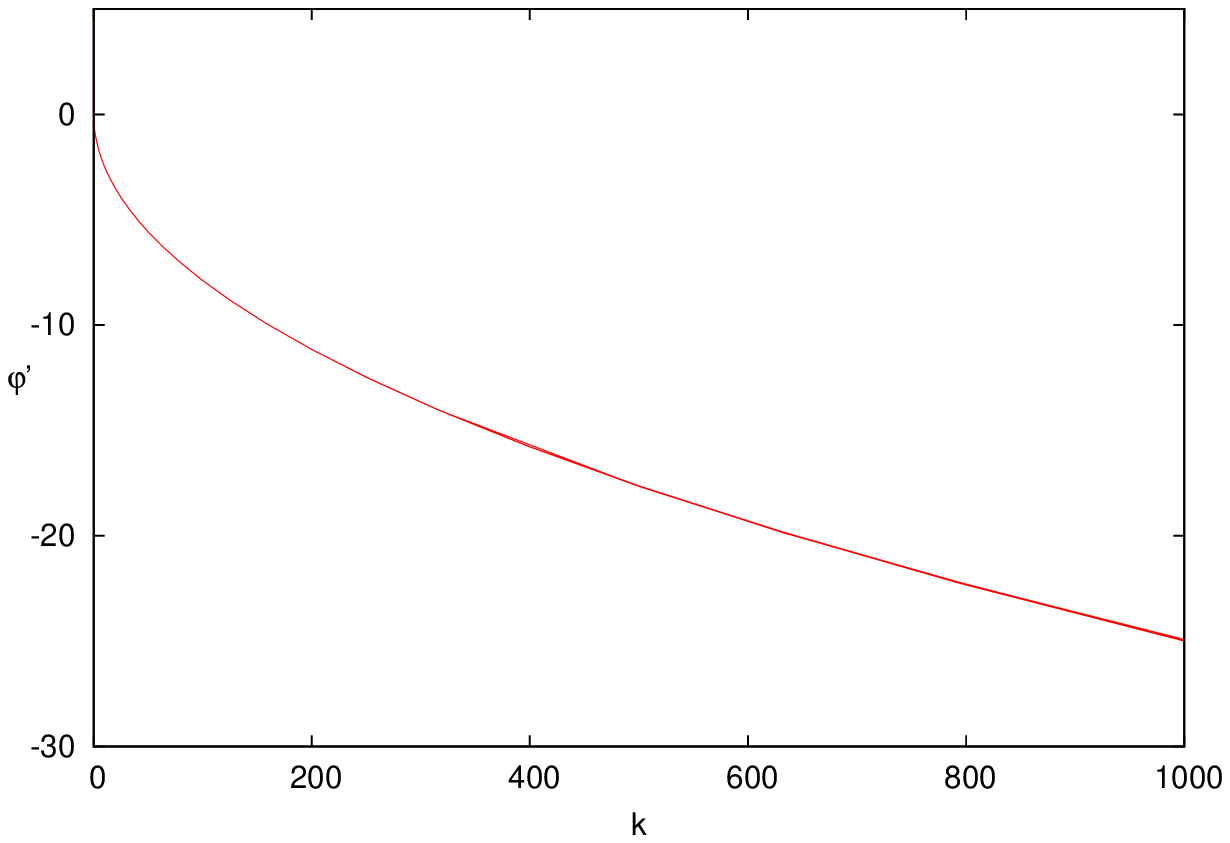}}
	\subfigure[]{\includegraphics[width=0.49\textwidth]{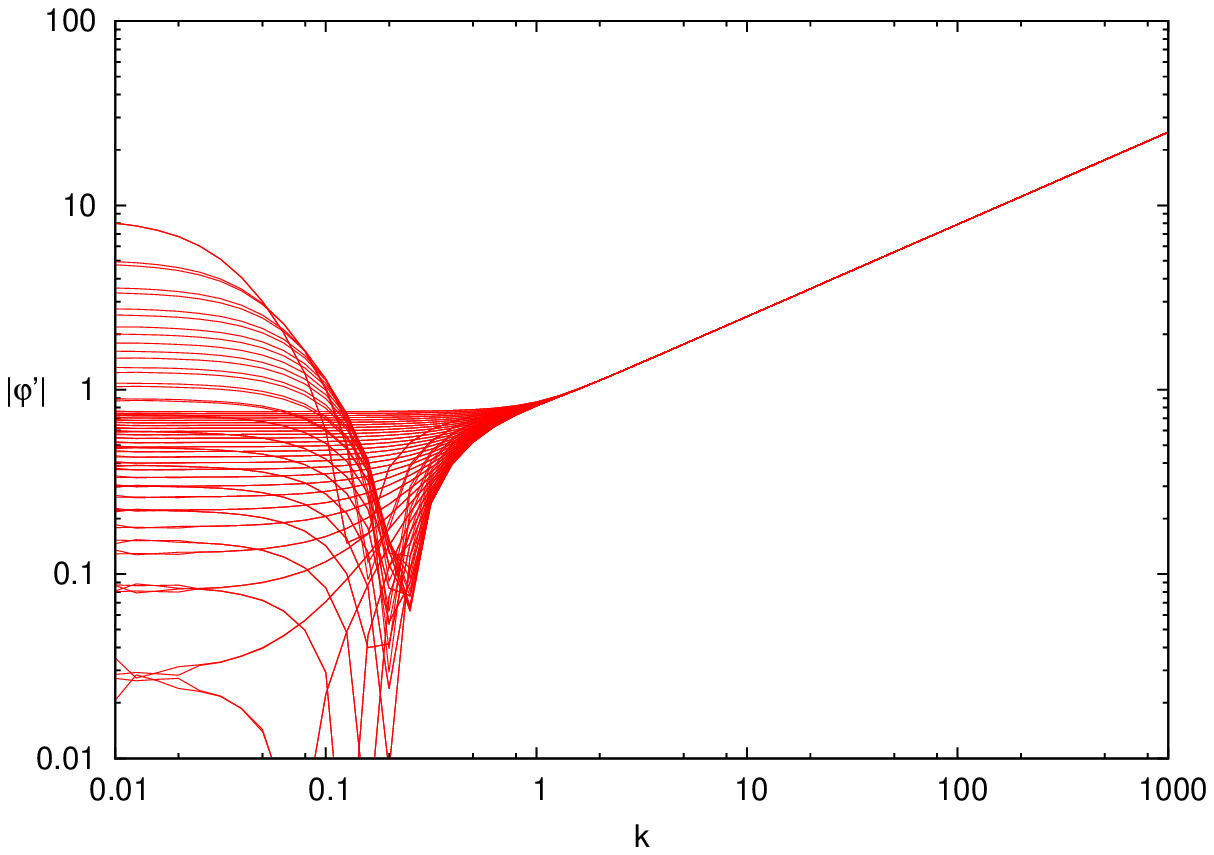}}
  \caption{The behavior of $\vp'(k)=\del_k\varphi(k)$ as a function of $k$ is 
  		presented for generic superselection sectors (without differentiating between 
  		branches corresponding to different values of $\epsilon$). $(a)$ presents its 
  		behavior in linear scale, whereas $(b)$ shows its absolute value in logarithmic 
  		scale.
	}\label{fig:dphase}
\end{figure*}

Unlike in the WDW case however, now we cannot determine $\partial_k \vp(k)$ 
analytically. In order to obtain a useful inequality we need to analyze the bounds 
on $\partial_k \vp(k)$ numerically. For that we implement the exact method used 
originally in \cite{kp-scatter} based on numerically probing the asymptotics of the 
function $\partial_k e_k(\nu)$. The only difference is that here, instead of using 
the original auxiliary basis elements \eqref{eq:aux-basis}, we use the corrected ones 
\eqref{eq:lim-imp}, which provide faster convergence and higher precision. The 
results are presented on Fig.~\ref{fig:dphase}. We see an explicit convergence (at 
large $k$) to the function
\begin{equation}
  \del_k \varphi(k) = A\sqrt{k} + o(1),
\end{equation}
where the constant $A$ has been determined numerically to equal $A= -0.789\pm 0.005$.

The exact behavior of $\del_k\varphi(k)$ depends on the superselection sector labeled 
by $\epsilon$. Let us start with the sector $\epsilon=0$. In that case, 
one of important observations following from numerical studies is the property that
\begin{equation}\label{eq:d2scatter}
  |\sqrt{k}\partial_k^2 \varphi(k)| \leq A/2 , 
\end{equation}
which allows us to conclude (via a derivation analogous to that of Sec.~5A 
in \cite{kp-scatter})
\begin{equation}
  \langle\Psi|\Delta \partial_k\varphi|\Psi\rangle 
  \leq \frac{A}{2} \langle\Psi|\Delta \sqrt{k}|\Psi\rangle .
\end{equation}
As a consequence the triangle inequality \eqref{eq:LQC-triangle} implies the 
following one, 
\begin{equation}\label{eq:LQC-triangle-final}
  \lim_{\Ag\to\infty} \langle {\Psi} | \Delta\wh{x}_{\Ag} | {\Psi} \rangle  
  \leq 
  \lim_{\Ag\to-\infty} \langle {\Psi} | \Delta\wh{x}_{\Ag} | {\Psi} \rangle 
  + A \langle \Psi|\Delta \sqrt{k}|\Psi\rangle ,
\end{equation}
which only involves observables with a clear physical interpretation
as the observable $\sqrt{k}$ can be easily replaced by the Dirac 
observable $\sqrt{\Pg}$.

\begin{figure}[tbh!]
	\includegraphics[width=0.96\textwidth]{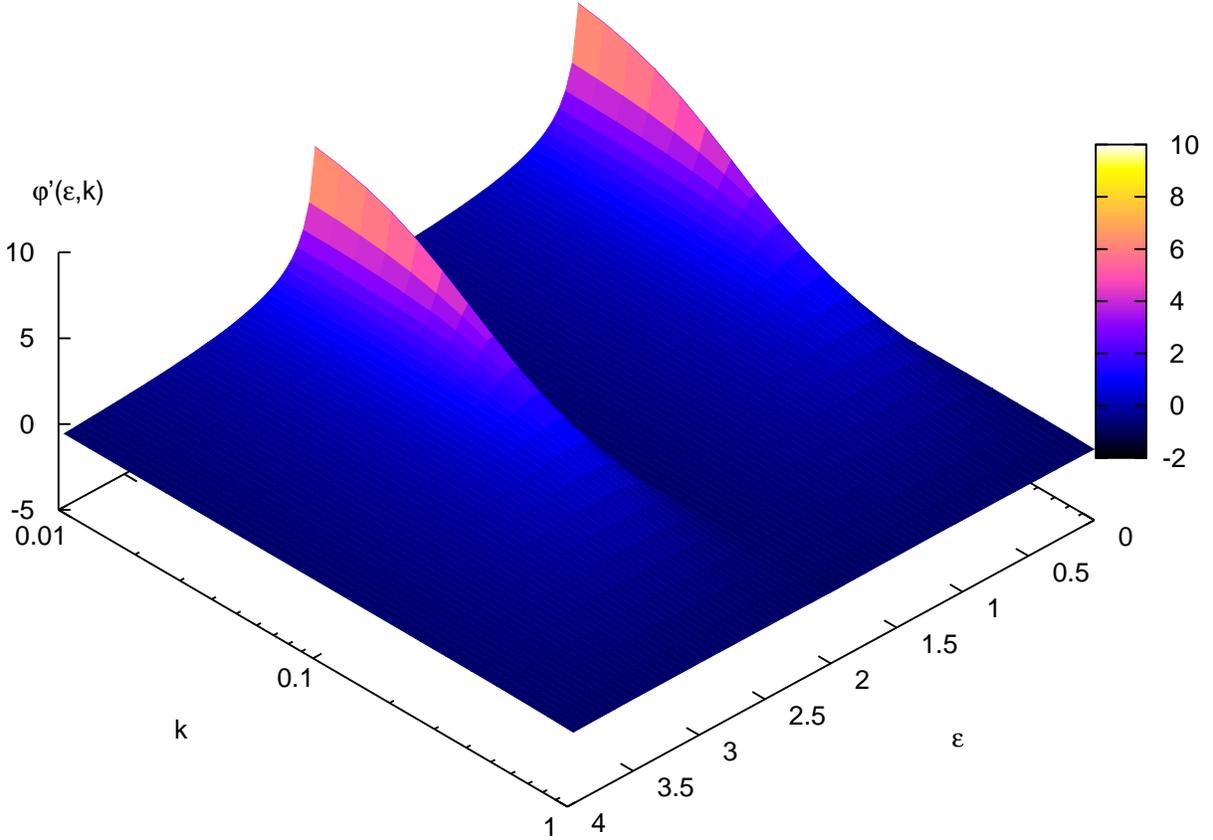}
	\caption{The behavior of $\varphi'=\del_k\vp$ as a function of both $k$ and 
		the superselection sector label $\epsilon$ is shown for small $k$.
	}\label{fig:dphi-small-3d}
\end{figure}

In the case of generic $\epsilon$ the situation is slightly more involved, as the 
numerical studies show significant differences in the behavior of $\del_k\varphi(k)$ for 
small values of $k$. We observe the right-hand discontinuity at $\epsilon=0$ and $\epsilon=2$. 
The bound \eqref{eq:d2scatter}, while preserved 
for $k > k_\star \approx 0.15$ may be violated for $k < k_\star$. The exact behavior
of $\del_k\varphi$ as the function of both $\epsilon$ and $k$ is presented on 
Fig.~\ref{fig:dphi-small-3d}.
As a consequence, for generic $\epsilon$ the triangle inequality \eqref{eq:LQC-triangle-final}
is ensured to hold strictly only for the states whose support does not overlap with 
the set $k\in[0,k_\star]$.


\raggedright

\end{document}